%% file: main.tex
\documentclass{article}
\pdfoutput=1
\usepackage{jcappub}
\usepackage{amsmath}
\usepackage{bm}
\usepackage{graphicx}
\usepackage{enumitem}
\usepackage{geometry}
\usepackage{xspace}
\usepackage{pdflscape}
\usepackage{dsfont}
\usepackage{MnSymbol}
\usepackage{tcolorbox}

\usepackage{bbm}

\usepackage[compat=1.1.0]{tikz-feynman}

\usepackage{mathtools}

\usepackage{relsize}

\allowdisplaybreaks

\makeatletter
\gdef\@fpheader{}
\g@addto@macro\bfseries{\boldmath}
\makeatother

\input{newcommands}

\newcommand{\deflen}[2]{%
    \expandafter\newlength\csname #1\endcsname
    \expandafter\setlength\csname #1\endcsname{#2}%
}

\usepackage{caption}
\usepackage{subcaption}
\usepackage{mathrsfs}
\usepackage{graphicx}
\newcommand*{\Scale}[2][4]{\scalebox{#1}{$#2$}}
\newcommand{\exd}{{ \mathrm{d}} }
\def\bea{\begin{eqnarray}}
\def\eea{\end{eqnarray}}

\def\ssA{{\scriptscriptstyle A}}
\def\ssB{{\scriptscriptstyle B}}

\def\mfs{{\mathfrak{s}}}
\def\mfr{{\mathfrak{r}}}

\def\smath#1{\text{\scalebox{.85}{$#1$}}}
\def\sfrac#1#2{\smath{\frac{#1}{#2}}}
\def\mpl{M_{\rm Pl}}

\subheader{}
 
\title{Decoherence out of fire: Purity loss in expanding and contracting universes}

\author[a,b,c]{Thomas Colas,}
\author[d,e,f]{Claudia de Rham,}
\author[d,e]{and Greg Kaplanek}

\affiliation[a]{Department of Applied Mathematics and Theoretical Physics, University of Cambridge, Wilberforce Road, Cambridge, CB3 0WA, UK}

\affiliation[b]{Universit\'e Paris-Saclay, CNRS, Institut d'Astrophysique Spatiale, 91405, Orsay, France}

\affiliation[c]{Laboratoire de Physique de l'\'Ecole Normale Sup\'erieure, ENS, Universit\'e PSL, CNRS, Sorbonne Universit\'e, Universit\'e Paris Cit\'e, F-75005 Paris, France}

\affiliation[d]{Department of Physics, Blackett Laboratory, Imperial College, London, SW7 2AZ, UK}

\affiliation[e]{Perimeter Institute for Theoretical Physics, Waterloo, Ontario, N2L 2Y5, Canada}

\affiliation[f]{CERCA, Department of Physics, Case Western Reserve University, 10900 Euclid Ave, Cleveland,
OH 44106, USA}

\emailAdd{tc683@cam.ac.uk}
\emailAdd{c.de-rham@imperial.ac.uk}
\emailAdd{g.kaplanek@imperial.ac.uk}

\date{today}

\begin{document}

\sloppy

\abstract{We investigate quantum decoherence in a class of models which interpolates between expanding (inflation) and contracting (ekpyrosis) scenarios. For the cases which result in a scale-invariant power spectrum, we find that ekpyrotic universes lead to complete decoherence of the curvature perturbation before the bounce. This is in stark contrast to the inflationary case, where recoherence has been previously observed in some situations. Although the purity can be computed for couplings of all sizes, we also study the purity perturbatively and observe that late-time (secular growth) breakdown of perturbation theory often occurs in these cases. Instead, we establish a simple yet powerful late-time purity resummation which captures the exact evolution to a remarkable level, while maintaining analytical control. We conclude that the cosmological background plays a crucial role in the decoupling of the heavy fields during inflation and alternatives.}


\maketitle

\section{Introduction}
\label{sec:intro}

The amplification of vacuum quantum fluctuations in the presence of a dynamical background is one of the cornerstones of primordial cosmology. It provides a mechanism which inflation \cite{Starobinsky:1979ty,Guth:1980zm, Starobinsky:1980te,Sato:1980yn, Linde:1981mu, Mukhanov:1981xt,  Mukhanov:1982nu, Guth:1982ec, Albrecht:1982wi, Starobinsky:1982ee, Hawking:1982cz, Linde:1983gd, Bardeen:1983qw, Mukhanov:1988jd} and many alternatives \cite{Alexander:2000xv, Khoury:2001bz, Khoury:2001wf, Martin:2001ue, Lyth:2001pf,Khoury:2001zk,Brandenberger:2001bs,Finelli:2001sr,Steinhardt:2002ih, Brandenberger:2001kj, Kallosh:2001ai, Peter:2002cn, Tsujikawa:2002qc, Kofman:2002cj, Notari:2002yc,Gasperini:2002bn,Gratton:2003pe,Tolley:2003nx,Geshnizjani:2005hc,Bozza:2005xs,Bozza:2005wn,Creminelli:2007aq,Buchbinder:2007ad,Lehners:2007ac,Koyama:2007ag,Tolley:2007nq,Noller:2009tj,Linde:2009mc,Lehners:2011kr,Peter:2006hx, Finelli:2007tr, Brandenberger:2016vhg, Agullo:2016tjh, Barrau:2013ula,Li:2013hga,Battarra:2013cha,Battefeld:2014uga,Ijjas:2015zma,Fertig:2016bki,Levy:2017ejc,Ijjas:2018qbo,Ijjas:2019pyf,Ijjas:2020cyh} 
rely on to generate cosmological inhomogeneities seeding the Cosmic Microwave Background (CMB) \cite{Aghanim:2018eyx,Planck:2018jri} and the Large Scale Structure (LSS) 
\cite{SDSS:2005xqv,BOSS:2014hwf,Colas:2019ret,DES:2022qpf} of the universe. The cosmological background is often compared to the role of the external electric field in the Schwinger effect \cite{Grishchuk:1990bj,Grishchuk:1992tw,Albrecht:1992kf, Polarski:1995jg, Lesgourgues:1996jc, Kiefer:1998qe, Martin:2007bw}, where the vacuum amplification here relies on the peculiar interplay between the classical geometry and quantum fluctuations living on the curved spacetime.\\

In this article, our goal is to investigate how different background evolution leading to the same observational features ({\it i.e.}~a nearly scale-invariant power spectrum for curvature perturbations compatible with current CMB and LSS constraints) might generate drastically distinct quantum dynamics for the fluctuations. Based on \cite{Creminelli:2007aq,Buchbinder:2007ad,Lehners:2007ac,Koyama:2007ag,Tolley:2007nq}, we consider a class of two-field models admitting scaling solutions which provides a framework interpolating between expanding (inflation) and contracting (ekpyrosis) scenarios.  \\

The question of quantum squeezing and decoherence in a similar two-field system of the ekpyrotic scenario was considered in \cite{Battarra:2013cha}, where it was shown that quantum decoherence occurs very efficiently once interactions between the entropic and adiabatic modes are switched on as one approaches the bounce. In this work we generalize these findings in multiple ways: first, the model we consider can be used to describe both a contracting ekpyrosis model or alternatively an inflationary expanding one, which enables a precise identification of the conditions under which decoherence occurs efficiently or fails to do so. Secondly, we consider the generation of a scale invariant power spectrum in the curvature mode at the same time as the interplay between the adiabatic and entropic modes, which in addition relies on a slightly different class of interactions between these modes. Finally, our findings lead to a resummed analytic estimation of the purity which matches the numerically integrated one with remarkable accuracy while allowing us to infer a deeper insight on the scaling and asymptotic behaviour of the purity.\\

To be more concrete, working at the level of the fluctuations, we consider the evolution of the curvature perturbation when the isocurvature modes have been integrated out. Using recent advances in the Open EFT and cosmological open-quantum-system programs \cite{breuerTheoryOpenQuantum2002, Koks:1996ga, Burgess:2006jn,Anastopoulos:2013zya, Fukuma:2013uxa, Burgess:2014eoa, Burgess:2015ajz, Boyanovsky:2015xoa, Boyanovsky:2015jen, Boyanovsky:2015tba, Nelson:2016kjm,Choudhury:2017bou,Choudhury:2017qyl,Hollowood:2017bil,Shandera:2017qkg,Boyanovsky:2018fxl,Boyanovsky:2018soy, Martin:2018zbe,Choudhury:2018ppd, Bohra:2019wxu, Akhtar:2019qdn, Kaplanek:2019dqu, Brahma:2020zpk, Kaplanek:2020iay, Rai:2020edx, Burgess:2021luo, Kaplanek:2021fnl, Brahma:2021mng, Banerjee:2021lqu, Oppenheim:2022xjr, Brahma:2022yxu, Kaplanek:2022xrr, Kaplanek:2022opa, Colas:2022hlq, Colas:2022kfu, DaddiHammou:2022itk, Burgess:2022nwu, Burgess:2022rdo, Cao:2022kjn, Prudhoe:2022pte, Raveendran:2022dtb, Raveendran:2023dst, Colas:2023wxa, Brahma:2023hki, Sharifian:2023jem, Alicki:2023tfz, Alicki:2023rfv, Ning:2023ybc}, we characterize the decoherence mechanism \cite{Zurek:1981xq,Zurek:1982ii,Joos:1984uk} which drives the quantum state of the curvature perturbations from a pure state to a mixed state due to the presence of the isocurvature modes. While the role of the entropic mass in the de(re)coherence mechanism has been highlighted in \cite{Colas:2022kfu}, we here emphasize the critical role of the background evolution in steering the quantum state of the system towards a mixed state. In the ekpyrotic scenario, we demonstrate that
curvature perturbations always decohere  provided the entropic sector is not so heavy that it entirely decouples. As the mass of the entropic sector increases, the time taken to decohere increases up to the point where decoherence happens at a time so close to the Big Crunch singularity that the EFT can no longer be trusted. Our main findings are summarized in Table \ref{tab:summary} which provides the late-time scaling of the purity of a mode $k$ as a tracer of the level of mixedness of the system's state. These results corroborate the findings of \cite{Raveendran:2023dst} which studies similar systems in the presence of sharp features (transient turns in the field space).\\

\begin{table}
    \centering
    \begin{tabular}{cccc}
         & Massless & Conformal & Heavy \\
        Expanding & $(-k \eta)^{-1}$ & $1/\log(-k \eta)$ & $1 - \ee^{-2 \pi m/H}$ \\
        Contracting & $(-k \eta)^{-3}$ & $(-k \eta)^{-2}$ & $(-k \eta)^{-3/2}$
    \end{tabular}
    \caption{Scaling of the late-time purity of the curvature perturbations for a given mode $k$ in terms of the conformal time $\eta$ for different isocurvature masses.}
    \label{tab:summary}
\end{table}

The rest of this work is structured as follows: In Section \S\ref{sec:Ekpy} we review the two-field model from \cite{Tolley:2007nq} which accommodates both expanding (inflation) and contracting (ekpyrosis) scenarios, with the goal to briefly summarize main results from the literature and to set up the formalism for later sections. In Section \S\ref{sec:EoM}, we derive the exact transport equations, which leads to an equation of motion for the purity corresponding to the curvature perturbation. We derive expressions for the purity in perturbation theory, as well as a late-time resummation that will prove insightful for later analysis. In Section \S\ref{sec:inflation} we consider first the expanding phase corresponding to usual slow-roll inflation, comparing the purity in the exact, perturbative and resummed descriptions. The results here closely follow \cite{Colas:2022kfu} which observes the phenomenon of purity freezing at late times. We then derive an analytic expression for the value of the asymptotic purity in the regime of large masses, where one approaches (but does not reach) the decoupling limit. In Section \S\ref{sec:contract}, we consider a contracting phase consistent with scale invariant power spectra and find that no recoherence ever occurs within the regime of validity of the EFT, irrespectively of the mass of the isocurvature mode (contrary to the inflationary case). We again compare the exact purity against approximations, finding our resummation useful as an analytic technique for quantifying how fast decoherence proceeds. Our conclusions are gathered in Section \S\ref{sec:conc} followed by two Appendices in which technical details are deferred. 

\vspace{2mm}

{\bf Conventions}: we use $\prime$ to denote a derivative with respect to conformal time $\eta$, while dots $\cdot$ to denote derivatives with respect to cosmic time $t$ (related by $a \exd \eta = \exd t$).

\section{Two-field primordial universe models}
\label{sec:Ekpy}

The study of cosmological inhomogeneities evolving in a symmetric background plays a crucial role in our current understanding of the primordial universe from pioneering works \cite{Starobinsky:1979ty,Guth:1980zm, Starobinsky:1980te,Sato:1980yn, Linde:1981mu, Mukhanov:1981xt,  Mukhanov:1982nu, Guth:1982ec, Albrecht:1982wi, Starobinsky:1982ee, Hawking:1982cz, Linde:1983gd, Bardeen:1983qw, Mukhanov:1988jd}
to latest observations \cite{Aghanim:2018eyx,Planck:2018jri,SDSS:2005xqv,BOSS:2014hwf,DES:2022qpf, Colas:2019ret, Cabass:2022wjy, DAmico:2022gki, Cabass:2022ymb, Cabass:2022epm, Krolewski:2023egq}. The dynamics of the background is imprinted in the evolution of the fluctuations which can be used as tracers of the geometry \cite{Chen:2012ja, Chen:2015lza, Chen:2018sce}. It is then crucial to determine which properties are inherited from the intrinsic properties of the fundamental constituents (mass, spin, chemical potential) and which ones are acquired dynamically from the background evolution. 

Inflation is the usual paradigm for explaining the source of the nearly scale-invariant, adiabatic primordial density fluctuations which seed large scale structure (LSS) formation as well as the signature seen in the cosmic microwave background (CMB) today. An alternative framework to inflation are the so-called bouncing or cyclic models of cosmology 
\cite{Alexander:2000xv, Khoury:2001bz, Khoury:2001wf, Martin:2001ue, Lyth:2001pf,Khoury:2001zk,Brandenberger:2001bs,Finelli:2001sr,Steinhardt:2002ih, Brandenberger:2001kj, Kallosh:2001ai, Durrer:2002jn, Peter:2002cn, Tsujikawa:2002qc, Kofman:2002cj, Notari:2002yc, Gasperini:2002bn,Gratton:2003pe,Tolley:2003nx,Geshnizjani:2005hc,Bozza:2005xs,Bozza:2005wn,Creminelli:2007aq,Buchbinder:2007ad,Lehners:2007ac,Koyama:2007ag,Tolley:2007nq,Noller:2009tj,Linde:2009mc,Lehners:2010fy,Lehners:2011kr,Peter:2006hx, Finelli:2007tr, Brandenberger:2016vhg, Agullo:2016tjh, Barrau:2013ula,Li:2013hga,Battarra:2013cha,Battefeld:2014uga,Ijjas:2015zma,Fertig:2016bki,Levy:2017ejc,Ijjas:2018qbo,Ijjas:2019pyf,Ijjas:2020cyh}, in which the universe undergoes a contracting phase {\it before} the hot Big Bang, creating the density perturbations imprinted in the LSS and CMB today. 

In the original ekpyrotic scenario, first introduced using a single field in \cite{Gratton:2003pe}, density perturbations are generated during a contacting phase in which the equation of state parameter satisfies $w > 1$. It was soon realized that the curvature perturbation $\zeta$ is only sensitive to the decaying mode (and not sensitive to the nearly scale-invariant growing mode of the Newtonian potential $\Phi$) -- the reason for this is that $\zeta$ and $\Phi$ correspond in a contracting phase to two physically distinct modes\footnote{Related by $\zeta = \frac{2}{3a^2(1+w)} \left( \frac{\Phi}{a'/a^3} \right)^{\prime}$.}, unlike in inflation. Using a single-field and a contracting phase therefore makes it difficult to generate the required scale-invariant spectrum for $\zeta$ (which in cyclic scenarios gives the perturbation amplitude at horizon re-entry after its spectrum is determined by the bounce), where one finds in the simplest setting that $\zeta$ has a blue tilt with $n_{s} \sim 3$. 

There have been various attempts at resolving this issue in the literature using a single scalar field (see {\it e.g.}~\cite{Peter:2006hx,Brandenberger:2020wha}), yet, a standard approach consists in considering instead a two-field model of the ekpyrotic scenario (for instance \cite{Finelli:2007tr}) which circumvents the above problem in the contracting phase. In this work, we analyze such a model which can be used to parameterize both a contracting universe as well as an expanding one so as to cover inflation as well. In this respect, it provides an efficient benchmark of the impact by the background on the dynamics of cosmological inhomogeneities in the inflation scenario and alternatives. We avoid any analysis of tensor perturbations in this work.

\subsection{Scaling solutions}

We follow the two-field formalism of \cite{Tolley:2007nq} which considers a mechanism for generating scale-invariant spectra in both expanding and contracting universes. The starting point is the action
\begin{equation} \label{actionstart}
S \ = \ \int \exd^4 x\; \sqrt{-g} \; \bigg[ \frac{1}{2} \mpl^2 R - \frac{1}{2} G_{ab} g^{\mu\nu} \partial_\mu \phi^a \partial_\nu \phi^b  - V( \boldsymbol{\phi} ) \bigg]\,,
\end{equation}
where two fields $\boldsymbol{\phi} = \{ \phi^1, \phi^2 \} $ couple to gravity, while also having their own metric on field space $G^{ab}$. We will be interested in {\it scaling solutions} in the above context where all contributions to the energy density scale the same way in time\footnote{This turns out to arise when there exists a homothetic Killing vector $K$ (satisfying $\nabla_a K_b - \nabla_b K_a = C g_{ab}$ for some constant $C$) which is also timelike.}. A scaling solution of this type turns out to exists when there is a continuous transformation with parameter $\lambda$ such that 
\begin{equation} \label{scalingsol}
\frac{\exd \phi^a}{\exd \lambda}  = K^a[\boldsymbol{\phi}] \  , \qquad g_{\mu\nu} \to e^{\lambda} g_{\mu\nu}  \ , \qquad S \to e^{\lambda} S \ , 
\end{equation}
where any such transformation preserves the equations of motion. These conditions imply that $K$ is a Killing vector on the field space and that the potential scales as $V \to e^{-\lambda} V$ under the above transformation.

One can pick a convenient basis for the scalar degrees of freedom made of the direction of the (field space) Killing vector to be in the $\phi$-direction, with the second field $\sigma$ pointing in the orthogonal direction. One can then use the coordinate freedom to put the kinetic term into the form\footnote{Note there is generically a cross-term of the form $A(\sigma) \; g^{\mu\nu}  \partial_{\mu} \phi \partial_{\nu} \sigma$ in the kinetic term above, but $A(\sigma)$ is only defined up to $A \to A + \partial_{\sigma} \alpha$ corresponding to redefinitions $\phi \to \phi + \alpha(\sigma)$ -- for the two-field case considered here one may use this freedom to set the cross-term to zero.}
\begin{equation} \label{kinetic}
 g^{\mu\nu} G_{ab}  \partial_\mu \phi^a \partial_\nu \phi^b \ = \  f(\sigma) \; g^{\mu\nu} \partial_{\mu} \phi \partial_{\nu} \phi \; + \; g^{\mu\nu} \partial_{\mu} \sigma \partial_{\nu} \sigma \,, 
\end{equation}
and the scaling behaviour forces the potential into the form
\begin{equation}
 V(\boldsymbol{\phi}) \ = \ V_0\; h(\sigma) \; e^{- \; \sqrt{2 \epsilon} \phi} \; \,, 
\end{equation}
for constant\footnote{Note that \cite{Tolley:2007nq} uses the notation $c = \sqrt{2\epsilon}$. We utilize the parameter $\epsilon$ since it will turn out to match the slow-roll parameter for the scale factor we will use later, keeping in mind here that we treat $\epsilon$ as constant.} $\epsilon$. One then rescales $\phi$ so that $f(0) = 1$ is satisfied and also rescales $V_0$ so that $h(0) = 1$, so that the background scaling solution is simply 
\begin{equation}
S \ = \ \int \exd^4 x\; \sqrt{-g}\; \bigg[ \frac{1}{2} \mpl^2 R - \frac{1}{2}  g^{\mu\nu}\partial_{\mu} \phi \partial_{\nu} \phi  - V_0 e^{- \sqrt{2\epsilon} \phi} \bigg] \  .
\end{equation}
Considering a background conformal FLRW metric with $\sigma = 0$, the scale factor $a$ and scalar field scale as\footnote{One can show that $\phi_0 = \frac{1}{\sqrt{2\epsilon}} \log \left( \frac{(\epsilon - 1)^2 V_0^2 \eta_0^2 }{ 3 - \epsilon } \right)$, where we note that $V_0 > 0$ for $\epsilon < 3$ and $V_0 <0$ for $\epsilon > 3$. In the special case that $\epsilon = 3$ (or $w=1$) the scalar field is free since the potential vanishes with $V_0 = 0 $ via the background equations and $\phi_0$ cannot be fixed in this case.} 
\begin{eqnarray}
a(\eta)  =  \left( \frac{\eta}{\eta_0} \right)^{\tfrac{1}{\epsilon - 1}} \label{scalefactor} \, \qquad {\rm and }\qquad
\phi(\eta)  =  \phi_0 + \frac{\sqrt{2\epsilon}}{\epsilon - 1} \log\left( \frac{\eta}{\eta_0} \right)\,,
\end{eqnarray}
where we work in terms of the conformal time $\eta$ (varying from $-\infty$ to $0^-$ during the pre-hot Big Bang era) and integration constant $\eta_0$.  Note furthermore that one may use this to express the equation of state parameter as
\begin{equation}
w = \frac{2}{3} \epsilon - 1 \ .
\end{equation}
Perturbing around the background with $\sigma \simeq 0 + \delta \sigma$ here, we Taylor expand the functions $f$ and $h$ appearing in the kinetic term (\ref{kinetic}) such that
\begin{eqnarray}
f(\sigma) \simeq 1 + f_1 \; \delta \sigma + \frac{1}{2} f_2 \; \delta \sigma^2 + \ldots \qquad \mathrm{and} \qquad h(\sigma) \simeq 1 + h_1 \; \delta \sigma + \frac{1}{2} h_2 \; \delta\sigma^2 + \ldots  \ .
\end{eqnarray}
Enforcing that $\sigma =0$ is consistent with the background equations of motion gives\footnote{ The apparent singularity at $\epsilon=3$ (or equivalently $w=1$) appearing in Eq.~(\ref{h1_f1}) really means that $f_1=0$ and that $h_1$ cannot be fixed by the background equations. This is a reflection of the fact that for $w=1$ the scalar field is free with $V_0 = 0$ (see footnote 5). }
\begin{equation} \label{h1_f1}
h_1 = \frac{\epsilon f_1}{3 - \epsilon} \ .
\end{equation}
This implies that the free parameters at this level are $(\epsilon,f_1, f_2,h_2)$. Expanding about the background $\phi \to \phi + \delta \phi$ in (\ref{scalefactor}) we treat the scalar fluctuation in the ADM formalism, so that the usual gauge-invariant variable is $\zeta = \mathcal{R} + \frac{a'}{a} \frac{\delta\phi}{\phi}$ with $\mathcal{R}$ the spatial metric perturbation. We choose to work in the uniform-field (or co-moving) gauge where $\delta \phi =0 $ so that the gauge-invariant variable $\zeta$ is exactly the metric perturbation. The fluctuation $\delta \sigma$ (or isocurvature mode) is already gauge-invariant since its background solution is $\sigma =0$.

Writing the action in terms of quantities $\zeta$ and $\delta\sigma$ results in the action
\begin{eqnarray} \label{2field_fluc}
S^{(2)} & = & \int \exd^3\bmk\; \exd\eta \; a^2(\eta) \; \bigg[ \epsilon \zeta'_{\bm{k}} \zeta'_{-\bm{k}} - \epsilon k^2 \zeta_{\bm{k}} \zeta_{-\bm{k}} + \frac{1}{2} \delta\sigma'_{\bm{k}} \delta\sigma'_{-\bm{k}} - \frac{1}{2} k^2 \delta\sigma_{\bm{k}} \delta\sigma_{-\bm{k}} \\
& \ & \;  - \; \frac{3 h_2 - \epsilon (f_2 + h_2)}{( \epsilon - 1)^2 \eta^2} \delta \sigma_{\bm{k}} \delta \sigma_{-\bm{k}} + \frac{2 \epsilon f_1}{(\epsilon - 1) \eta} \zeta'_{\bm{k}}  \delta \sigma_{-\bm{k}} \bigg]  \ .\notag
\end{eqnarray}
When $f_1 = 0$ the adiabatic and entropic modes are uncoupled. In what follows, we assume that $f_1 \neq 0$ which results in non-trivial dynamics of $\zeta$ and $\delta\sigma$. It is not possible to exactly solve for the mode functions in this case (despite the fact that the action (\ref{2field_fluc}) is truncated at Gaussian order -- see {\it e.g.}~\cite{An:2017hlx, Werth:2023pfl} for a discussion). We rely on a numerical approach to compute the exact dynamics \cite{Dias:2016rjq, Ronayne:2017qzn, Werth:2023pfl, Raveendran:2023dst, Pinol:2023oux}, as well as a perturbative approach in $f_1$ to glean some analytic control, before providing a better analytic method in the form of a resummation. 

\subsection{Expanding and contracting cases}
\label{sec:scaleinv}

We here briefly summarize some basic results for observables of the theory from \cite{Tolley:2007nq}, namely explore the value of the spectral index and the constraints that the requirement of scale invariance enforces on the free parameters $(\epsilon,f_1, f_2,h_2)$ of the theory. The power spectra are defined in the usual way in terms of two-point correlators of the theory where 
\begin{equation}  \label{PowerSpectrum_def}
\langle   \zeta_{\bm{k}}(\eta_{\mathrm{f}}) \zeta_{\bm{q}}(\eta_{\mathrm{f}}) \rangle =: \delta(\bm{k} + \bm{q}) \frac{2\pi^2}{k^3} P_{\zeta}(k) \qquad \mathrm{and} \qquad \langle \sigma_{\bm{k}}(\eta_{\mathrm{f}}) \sigma_{\bm{q}}(\eta_{\mathrm{f}}) \rangle =: \delta(\bm{k} + \bm{q}) \frac{2\pi^2}{k^3} P_{\sigma}(k)\,,
\end{equation}
where $\eta_{\mathrm{f}}$ is the time at which the generation process ends (such as at reheating in the inflationary/expanding case). The spectral index $n_s$ is then defined in terms of the scaling of the power spectrum with $k$ such that $P_{\zeta}(k) \propto k^{n_s - 1}$. 

Using the WKB approximation within the horizon and matching this solution at horizon-crossing with power-law super-horizon modes, the authors of \cite{Tolley:2007nq} identified scaling solutions giving rise to the {\it exact} spectral index
\begin{equation} \label{spectral}
n_{s} = \mathrm{min} \left\{ \; \frac{3 \epsilon - 1}{\epsilon - 1} , \; \frac{5 \epsilon - 7}{\epsilon - 1} , \; 4 - \frac{\sqrt{ (\epsilon - 3)^2 - 8 \epsilon \Delta }}{|\epsilon - 1|} \; \right\} \qquad \mathrm{with} \ \Delta := f_1^2 - \frac{1}{2} f_2 - \frac{1}{2} \left( 1 - \frac{3}{\epsilon} \right) h_2\ .
\end{equation}
Scale invariance is achieved for $n_{s} = 1$, and so as outlined in \cite{Tolley:2007nq} there are three different ways for scale invariance to be achieved in this model (depending on which term in (\ref{spectral}) dominates):

\vspace{2mm}

{\bf Case (i):} When $\epsilon = 0$ and so $w = -1$ makes the first term of (\ref{spectral}) dominate -- this is the usual expanding, nearly de Sitter inflationary universe where the linear action in this case coincides with that of a constant turn in field space \cite{Achucarro:2012sm, Cespedes:2012hu} studied first in the gelaton scenario \cite{Tolley:2009fg} and later in quasi-single field inflation \cite{Chen:2009zp, Assassi:2013gxa}. Note that decoherence in this class of models has been studied in the past (see {\it e.g.}~\cite{Prokopec:2006fc} for an early discussion), including more recently in \cite{Colas:2022kfu}. We use the same formalism here and so especially the discussion of the expanding phase closely follows and complements this work.

\vspace{1mm}

{\bf Case (ii):} When $\epsilon = \frac{3}{2}$ and so $w = 0$ makes the second term of (\ref{spectral}) dominate, provided that $\Delta > 0$. This gives rise to a contracting universe.

\vspace{1mm}

{\bf Case (iii):} When $\epsilon > \frac{3}{2}$ and so $w > 0$ makes the final term of (\ref{spectral}) dominate, provided that one also has $\Delta = \frac{3}{2} - \epsilon$. This again yields a contracting universe.

\vspace{2mm}

Notice that the mass term of $\delta \sigma$ in the action (\ref{2field_fluc}) has a time-dependent mass parameter which can be written in terms of $\Delta$ as 
\begin{equation} \label{eff_mass_sigma}
M_{\mathrm{eff}}^2(\eta) \ : = \ \frac{6 h_2 - 2 \epsilon (f_2 + h_2)}{( \epsilon - 1)^2\eta^2} \ = \ \frac{4 \epsilon (\Delta - f_1^2)}{( \epsilon - 1)^2\eta^2} \ .
\end{equation}
This means that in the case (ii), the effective mass squared $M_{\mathrm{eff}}^2$ can take any sign by varying $\Delta >0$ and the size of the coupling $f_1$. In case (iii), we instead require $\Delta = \frac{3}{2} - \epsilon < 0$ and so in this case $M_{\mathrm{eff}}^2$ is always negative.

The inflationary case has been shown to be always stable. The analysis in \cite{Tolley:2007nq} suggests that the background is stable for contracting solutions only when $\Delta > 0$ and $w>1$ -- from the above, we conclude that {\it none} of the contracting backgrounds which give scale-invariant spectra are stable. This implies that some degree of fine-tuning is required for the initial conditions of our contracting background solutions, a consideration we choose to overlook in this work. Instead, our focus lies on investigating the phenomenological consequences of the contracting cases (ii) and (iii). One of our motivations is to explore the impact of the background evolution on the quantum information properties of observable degrees of freedom, without necessarily advocating for the phenomenological feasibility of the model.

\subsection{Contracting universe and regime of validity}
\label{sec:cutoff}

The geometry of contracting phases also implies the existence of a Big Crunch at very late times, analogous to the necessity of a Big Bang at early times in an expanding universe. In the same sense that effective physical descriptions of inflation breakdown at very early times (times near the Big Bang), contracting universes have a time {\it after} which the contraction has proceeded to such a degree that we need new physics to properly describe what happens to the background.

We here estimate a cutoff time $\eta_{\mathrm{cut}}$ after which the physics breaks down in $\epsilon \geq \frac{3}{2}$ contracting phases. To proceed we note that the scale factor (\ref{scalefactor}) implies that the Hubble rate is
\begin{eqnarray}
H(\eta) \ := \ \frac{\dot{a}}{a} \ = \ \frac{a'}{a^2} \ = \ \frac{1}{(1-\epsilon) (- \eta_0)} \left( \frac{\eta}{\eta_0} \right)^{\tfrac{1}{1 - \epsilon } -1} \,,
\end{eqnarray}
which we see is negative for any $\epsilon \geq \frac{3}{2}$. The usual definition of the first slow-roll parameter is
\begin{eqnarray}
\epsilon_{H}(\eta) \ := \ - \frac{\dot{H}}{H^2} \ = \ \frac{H'}{aH^2} \ = \ \epsilon
\end{eqnarray}
which shows that for any phase this is a constant. Notice that the Hubble rate is constant only for the case of inflation with $\epsilon \ll 1$, in which case we identify $H = - \eta_0^{-1}$ -- otherwise the reference time $\eta_0$ does not have a clean interpretation on its own.

Keeping a conservative approach, a breakdown of the gravitational effective field theory will take place at the very least when curvature invariants become close to the Planck scale (likely well before). This implies a gravitational regime of validity where $|H(\eta)| \ll \mpl$ or more explicitly where $\eta \ll \eta_{\mathrm{cut}} \sim \eta_0 ( - \mpl \eta_0 )^{{1}/{\epsilon} - 1} $. Bearing in mind that in principle, the effective field theory we are using to describe this system could still break down well before the cutoff time $\eta_{\rm cut}$. However as we shall see, the loss of purity is so efficient in the contracting phase that the precise value of the cutoff time is not particularly relevant to our analysis, so long as there is a regime where the gravitational effective field theory is under control.

\section{Exact and perturbative treatments}
\label{sec:EoM}

When a quantum system is embedded in a wider \textit{environment}, the increase of information shared between the two sectors tends to drive the system's state toward a mixed state through the mechanism of \textit{quantum decoherence} \cite{Zurek:1981xq,Zurek:1982ii,Joos:1984uk}. In cosmology, the presence of primordial degrees of freedom decaying at late-time so that they are effectively unobservable provides a natural class of cosmological environments \cite{breuerTheoryOpenQuantum2002, Koks:1996ga, Burgess:2006jn,Anastopoulos:2013zya, Fukuma:2013uxa, Burgess:2014eoa, Burgess:2015ajz, Boyanovsky:2015xoa, Boyanovsky:2015jen, Boyanovsky:2015tba, Nelson:2016kjm, Hollowood:2017bil,Shandera:2017qkg,Boyanovsky:2018fxl,Boyanovsky:2018soy, Martin:2018zbe, Bohra:2019wxu, Akhtar:2019qdn, Kaplanek:2019dqu, Brahma:2020zpk, Kaplanek:2020iay, Rai:2020edx, Burgess:2021luo, Kaplanek:2021fnl, Brahma:2021mng, Banerjee:2021lqu, Oppenheim:2022xjr, Brahma:2022yxu, Kaplanek:2022xrr, Kaplanek:2022opa, Colas:2022hlq, Colas:2022kfu, DaddiHammou:2022itk, Burgess:2022nwu, Burgess:2022rdo, Cao:2022kjn, Prudhoe:2022pte, Colas:2023wxa, Brahma:2023hki, Sharifian:2023jem, Alicki:2023tfz, Alicki:2023rfv, Ning:2023ybc}. 

In this section, we derive the time evolution of the reduced density matrix $\varrho_{\ssA}$ of the \textit{open system} sector (the adiabatic modes) in the presence of an environment consisting of the isocurvature modes. The theory is Gaussian and so the dynamics is exactly solvable in principle, although in practice this needs to be done numerically for generic sized couplings between the adiabatic and isocurvature modes. We use the purity $\gamma := \mathrm{Tr}[\varrho^2_{\ssA}]$ as a measure of mixedness of the state which characterizes the decoherence undergone by the adiabatic mode -- we here derive an exact equation of motion for $\gamma$ (to be solved numerically in later sections), as well as an analogous expression in perturbation theory. We close off this section by motivating a resummed expression for the purity, which turns out to be convenient for later use in \S\ref{sec:inflation} and \S\ref{sec:contract}.

\subsection{Canonical variables, Hamiltonian and quantum evolution}

For the rest of the analysis, it will be convenient to work in terms of the canonical variables
\begin{eqnarray}
v := \sqrt{2 \epsilon} a M_{\mathrm{Pl}} \zeta \qquad \qquad \mathrm{and} \qquad \qquad \chi := a \delta \sigma \,,
\end{eqnarray}
which turns the above action (\ref{2field_fluc}) into
\begin{eqnarray}
S^{(2)} & = & \int \exd^3\bmk\; \exd\eta \; \bigg[ \frac{1}{2} \bigg( v'_{\bm{k}} - \frac{a'}{a} v_{\bm{k}} \bigg) \bigg( v'_{-\bm{k}} - \frac{a'}{a} v_{-\bm{k}} \bigg) - \frac{1}{2} k^2 v_{\bm{k}} v_{-\bm{k}}  \\
& \ & \; + \; \frac{1}{2} \bigg(\chi'_{\bm{k}} - \frac{a'}{a} \chi_{\bm{k}} \bigg)\bigg(\chi'_{-\bm{k}} - \frac{a'}{a} \chi_{-\bm{k}} \bigg)  - \frac{1}{2} \left[ k^2 + M_{\mathrm{eff}}^2(\eta) \right] \chi_{\bm{k}} \chi_{-\bm{k}} + \lambda(\eta) \bigg( v'_{\bm{k}} - \frac{a'}{a} v_{\bm{k}} \bigg) \chi_{-\bm{k}} \bigg]  \,, \notag
\end{eqnarray}
with $M_{\mathrm{eff}}^2(\eta)$ defined in (\ref{eff_mass_sigma}) and using the shorthand for the time-dependent coupling
\begin{eqnarray} \label{coupling_def}
\lambda(\eta) & : = & \frac{\sqrt{2\epsilon} f_1 }{ (\epsilon - 1) \eta } \ . 
\end{eqnarray}
Defining the respective canonical momenta
\begin{eqnarray}
p : = \frac{\partial L}{\partial v'}  =  v' - \frac{a'}{a} v + \lambda \chi \qquad \quad \mathrm{and}  \quad \qquad \pi : =  \frac{\partial L}{\partial \chi'} = \chi' - \frac{a'}{a} \chi\,,
\end{eqnarray}
and building the Hamiltonian with a Legendre transform as usual gives 
\begin{eqnarray} 
H(\eta) & = & \int \exd^3\bmk\; \exd\eta \; \bigg[ \frac{1}{2} p_{\bmk} p_{-\bmk}  + \frac{a'}{a} p_{\bmk} v_{-\bmk} + \frac{1}{2} k^2 v_{\bm{k}} v_{-\bm{k}} \\
& \ & \; + \; \frac{1}{2} \pi_{\bm{k}} \pi_{-\bm{k}} + \frac{a'}{a} \pi_{\bmk} \chi_{-\bmk}  + \frac{1}{2} \left[ k^2 + M_{\mathrm{eff}}^2(\eta) + \lambda^2(\eta) \right] \chi_{\bm{k}} \chi_{-\bm{k}} - \lambda(\eta) p_{\bmk} \chi_{-\bmk} \bigg] \,, \notag
\end{eqnarray}
where we note that
\begin{equation} \label{full_tdepmass}
M_{\mathrm{eff}}^2(\eta) + \lambda^2(\eta) \ = \ \frac{2 \epsilon (2\Delta - f_1^2)}{( \epsilon - 1)^2\eta^2} \,,
\end{equation}
which gives the value of the full time-dependent mass of the isocurvature mode. The phase space variables form the vector
\begin{equation}
\mathbf{z}_{\bm{k}} \ = \ \left( \; v_{\bm{k}}, \; p_{\bm{k}} , \; \chi_{\bm{k}}, \; \pi_{\bm{k}} \;  \right)\,,
\end{equation}
and for convenience we define the real and imaginary parts of the above fields as,
\begin{equation} \label{RI_label_def}
\mathbf{z}^{\mathrm{R}}_{\bm{k}} := \frac{\mathbf{z}_{\bm{k}} + \mathbf{z}_{-\bm{k}}}{\sqrt{2}} \qquad \mathrm{and} \qquad \mathbf{z}^{\mathrm{I}}_{\bm{k}} := \frac{\mathbf{z}_{\bm{k}} - \mathbf{z}_{-\bm{k}}}{\sqrt{2} i} \ .
\end{equation}
The density matrix is easily decomposed according to this basis and the Hamiltonian can be compactly written in the form\footnote{We integrate Fourier modes over the half-space $\mathbb{R}_{3}^{+}$ here to avoid double-counting degrees of freedom, since the properties $\mathbf{z}^{\mathrm{R}}_{-\bm{k}} = \mathbf{z}^{\mathrm{R}}_{\bm{k}}$ and $\mathbf{z}^{\mathrm{I}}_{-\bm{k}} =  -\mathbf{z}^{\mathrm{I}}_{\bm{k}}$ are used in deriving (\ref{H_def}) -- {\it i.e.}~using this symmetry causes cross-terms between $\mathrm{R}$ and $\mathrm{I}$ to cancel. See \cite{Martin:2018zbe} for more details.}
\begin{eqnarray} \label{H_def}
H(\eta) & = & \sum_{\mfs = \mathrm{R}, \mathrm{I} } \int_{\mathbb{R}^{3+} } \exd^3 \bm{k} \; \frac{1}{2} \; \mathbf{z}^{\mfs\; T}_{\bm{k}}(\eta) \mathbb{H}_{\bm{k}}(\eta) \mathbf{z}^{\mfs}_{\bm{k}}(\eta) \,,
\end{eqnarray}
where $\mathbb{H}_{\bm{k}}(\eta)$ is the $4\times 4$ matrix with
\begin{equation} \label{H_matrix}
\mathbb{H}_{\bm{k}}(\eta) := \left[ \begin{matrix} \mathbb{A}_{\bm{k}}(\eta) & \mathbb{V}(\eta) \\ \mathbb{V}(\eta)^{T} & \mathbb{B}_{\bm{k}}(\eta) \end{matrix}  \right] \,,
\end{equation}
where the block matrices in the above are defined by
\begin{equation}
\mathbb{A}_{\bm{k}}(\eta) := \left[ \begin{matrix} k^2 & \frac{a'(\eta)}{a(\eta)} \\
\frac{a'(\eta)}{a(\eta)} & 1 \end{matrix}  \right] \ , \quad  \mathbb{B}_{\bm{k}}(\eta) := \left[ \begin{matrix} k^2 + M_{\mathrm{eff}}^2(\eta) + \lambda^2(\eta) & \frac{a'(\eta)}{a(\eta)} \\
\frac{a'(\eta)}{a(\eta)} & 1 \end{matrix}  \right] \ , \quad  \mathbb{V}(\eta) := \left[ \begin{matrix} 0 & 0 \\
- \lambda(\eta) & 0 \end{matrix}  \right] \ .
\end{equation}
This completes all the required elements to quantize and evolve the system. The mixing coupling $\lambda$ that enters in $\mathbb{V}$ is the crucial ingredient that  enables decoherence to happen (although does not necessarily guarantee it on its own). In the analysis below, we shall use a combination of exact and perturbative analysis in $\lambda$.

\subsubsection{Quantization and state evolution}

Upon quantization, the field variables are promoted to operators obeying the canonical commutation relations
\begin{eqnarray}
{[} v_{\bm{k}}^{\mfs}(\eta) , p_{\bm{q}}^{\mfr}(\eta) ] = i \delta_{\mfs\mfr} \delta(\bm{k} - \bm{q}) \qquad \mathrm{and} \qquad {[} \chi^{\mfs}_{\bm{k}}(\eta) , \pi^{\mfr}_{\bm{k}}(\eta) ] = i \delta_{\mfs\mfr} \delta(\bm{k} - \bm{q})\,,
\end{eqnarray}
with all other commutators vanishing. The adiabatic mode is here treated as the open system living in sector $A$, and the environment is taken to be the isocurvature mode in sector $B$ -- the total Hilbert space is then a tensor product $\mathcal{H}_{\ssA} \otimes \mathcal{H}_{\ssB}$ of the system and environment Fock spaces (respectively), which means the Hamiltonian (\ref{H_def}) can be taken to decompose as
\begin{eqnarray} \label{H_split}
H(\eta) = H_{\ssA}(\eta) \otimes \mathcal{I}_{\ssB} + \mathcal{I}_{\ssA} \otimes H_{\ssB}(\eta) + H_{\mathrm{int}}(\eta)\,,
\end{eqnarray}
with $\mathcal{I}_{\ssA,\ssB}$ identities on each Hilbert space, and with the definitions:
\begin{eqnarray} 
H_{\ssA}(\eta) & := & \sum_{\mfs = \mathrm{R}, \mathrm{I} } \int_{\mathbb{R}^{3+}} \exd^3 \bm{k} \; \bigg( \frac{1}{2} \big[ p^{\mfs}_{\bm{k}}(\eta) \big]^2 + \frac{a'}{2a} \{ p_{\bm{k}}^{\mfs}(\eta) , v_{\bm{k}}^{\mfs}(\eta) \} +  \frac{1}{2} k^2 \big[ v^{\mfs}_{\bm{k}}(\eta) \big]^2 \bigg) \label{HA_def} \\ 
H_{\ssB}(\eta) & := & \sum_{\mfs = \mathrm{R}, \mathrm{I} } \int_{\mathbb{R}^{3+}} \exd^3 \bm{k} \; \bigg( \frac{1}{2} \big[ \pi^{\mfs}_{\bm{k}}(\eta) \big]^2 + \frac{a'}{2a} \{ \pi_{\bm{k}}^{\mfs}(\eta) , \chi_{\bm{k}}^{\mfs}(\eta) \} +  \frac{1}{2} \big( k^2 + M_{\mathrm{eff}}^2(\eta) \big) \big[ \chi^{\mfs}_{\bm{k}}(\eta) \big]^2 \bigg) \label{HB_def} \qquad \\ 
H_{\mathrm{int}}(\eta) & := &  \sum_{\mfs = \mathrm{R}, \mathrm{I} } \int_{\mathbb{R}^{3+}} \exd^3 \bm{k} \; \bigg( \; \lambda(\eta) \; p^{\mfs}_{\bm{k}}(\eta) \otimes \chi^{\mfs}_{\bm{k}}(\eta) + \lambda^2(\eta) \; \mathcal{I}_{\ssA} \otimes \big[ \chi^{\mfs}_{\bm{k}}(\eta) \big]^2 \bigg)\,.  \label{Hint_def}
\end{eqnarray}
Written in the interaction picture, the splitting of the Hamiltonian in (\ref{H_split}) is useful for perturbation theory later where we eventually will perturb in the coupling $\lambda$, (by assuming that the dimensionless coefficient $\frac{\sqrt{2\epsilon} f_1}{\epsilon-1}$ is small). 

The full quantum state $\rho(\eta)$ of both the system and its environment evolves in the interaction picture according to the Von Neumann equation such that
\begin{eqnarray} \label{vN_EQ}
\frac{\partial \rho}{\partial \eta} \ = \ - i [ H_{\mathrm{int}}(\eta) , \rho(\eta) ] \ .
\end{eqnarray}
The decomposition of the Hamiltonian (\ref{H_def}) according to modes $\bm{k} \in \mathbb{R}^{3+}$ and label $\mathfrak{s} = \mathrm{R}, \mathrm{I}$ is here useful, since one can show that the density matrix factorizes in the sense that
\begin{eqnarray} \label{rho_decomp}
\rho(\eta) \ = \ \bigotimes_{\bm{k},\mfs} \; \rho_{\bm{k}}^{\mfs}(\eta)\,,
\end{eqnarray}
where each degree of freedom $\rho_{\bm{k}}^{\mfs}(\eta)$ evolves independently according to the equation
\begin{eqnarray} \label{vN_mode}
\frac{\partial \rho_{\bm{k}}^{\mfs} }{\partial \eta} \ = \ - i \; \Big[ \lambda(\eta) \; p^{\mfs}_{\bm{k}}(\eta) \otimes \chi^{\mfs}_{\bm{k}}(\eta) + \lambda^2(\eta) \; \mathcal{I}_{\ssA} \otimes \big[ \chi^{\mfs}_{\bm{k}}(\eta) \big]^2 , \; \rho_{\bm{k}}^{\mfs}(\eta) \Big] \times \frac{(2\pi)^3}{\mathcal{V}}  \,.
\end{eqnarray}
This is the case here only because the theory considered here is Gaussian \cite{Martin:2018zbe,Grain:2019vnq,Colas:2021llj}, and where
\begin{eqnarray}
\delta^{(3)}(\mathbf{0}) = \frac{\mathcal{V}}{(2\pi)^3}\,,
\end{eqnarray}
is a volume factor included for dimensional reasons (see \cite{Burgess:2022nwu}). The initial conditions given are that of the Bunch-Davies state in the distant past where
\begin{equation} \label{BD}
\rho(\eta_{\mathrm{in}}) \ = \ | \mathrm{BD} \rangle \langle \mathrm{BD} |\,,  
\end{equation}
where we take $\eta_{\mathrm{in}} \to -\infty - i \varepsilon$ in calculations performed (with the $i \varepsilon$-prescription needed as always so as to project onto the interacting vacuum). In practice this means enforcing that canonical mode functions take a WKB form $\sim \frac{1}{\sqrt{2k}} (- k \eta)^{i s } e^{- i k \eta}$ on sub-horizon scales with $- k \eta \gg 1$ (see \cite{Tolley:2007nq}). 

\subsubsection{Reduced dynamics for the open system}

Our interest in this work is to quantify the amount of decoherence in the open system at late times given the initial condition (\ref{BD}). Tracking the evolution of sector $A$ then corresponds to computing the reduced density matrix
\begin{eqnarray}
\varrho_{\ssA\bm{k}}^{\mfs}(\eta) \ := \ \underset{B}{\mathrm{Tr}}\big[ \rho_{\bm{k}}^{\mfs}(\eta) \big]\,,
\end{eqnarray}
where the partial trace is performed over the environment degrees of freedom in sector $B$ (here the Fock space corresponding to each label $\bm{k}$ and $\mfs$). The resulting operator $\varrho_{\ssA}$ fully describes the open system in the sense that one can use it to express expectation values of operators $O_{\ssA}$ in the system sector with $\mathrm{Tr}_{\ssA}[ O_{\ssA} \varrho_{\ssA} ]$. Typically the open system sector is taken to be made out of the degrees of freedom whose observables are able to be measured directly, as is the case for the adiabatic mode (through its imprint on the CMB for example) -- this is what motivates our choice of the adiabatic mode being the open system in this work. 

In general, decoherence is the open system's tendency to lose quantum coherence \cite{Zurek:1981xq,Zurek:1982ii,Joos:1984uk} as the system leaks information into the environment. As the system decoheres, its reduced density matrix generically transitions from a pure state to a mixed state (and so the system is sometimes said to have classicalized as it is describable by classical statistical ensembles). Quantum decoherence of the adiabatic fluctuations might obstruct our ability to extract genuine quantum correlations at late-time \cite{Campo:2005sv, Maldacena:2015bha, Choudhury:2016cso, Choudhury:2016pfr,  Martin:2017zxs}. It might also be a desirable feature in the context of the so-called quantum-to-classical transition of cosmological inhomogeneities to explain the appearance of classical distributions \cite{Brandenberger:1990bx, Brandenberger:1992sr, Barvinsky:1998cq, Lombardo:2005iz, Kiefer:2006je, Martineau:2006ki, Burgess:2006jn, Prokopec:2006fc, Kiefer:2008ku, Nelson:2016kjm, Hollowood:2017bil, Martin:2018zbe, Martin:2018lin, Kanno:2020usf, Martin:2021znx, DaddiHammou:2022itk, Burgess:2022nwu}. Hence, we want to assess if the curvature perturbations have undergone decoherence at the dawn of the hot Big Bang. 

For single-field Gaussian systems, quantum decoherence tracers are fully determined by the so-called {\it purity} parameter \cite{Serafini:2003ke, Martin:2021znx, Colas:2021llj, Martin:2022kph},
\begin{eqnarray} \label{purity_def}
\gamma_{\bm{k}}^{\mfs}(\eta) \ : = \ \underset{A}{\mathrm{Tr}}\left[ \big( \varrho_{\ssA\bm{k}}^{\mfs}(\eta)\big)^2 \right]\,,
\end{eqnarray}
where the relation $0 \leq \gamma_{\bm{k}}^{\mfs} \leq 1$ always holds, taking the value $1$ only for pure states, or the value $0$ for maximally mixed states\footnote{In the infinite dimensional case considered here.}. Being a trace, this quantity is basis-independent and is invariant under field redefinitions (as long as the redefinition does not affect the bipartition \cite{Serafini:2003ke}).

\subsection{Exact evolution via transport equations}

In principle one may solve for $\rho_{\bm{k}}^{\mfs}$ in the von Neumann Eq.~(\ref{vN_mode}), and perform the required traces to compute $\gamma_{\bm{k}}^{\mfs}$ from Eq.~(\ref{purity_def}). In practice this proves to be cumbersome, and so we here use the {\it transport equation} approach \cite{Dias:2016rjq, Ronayne:2017qzn, Werth:2023pfl, Raveendran:2023dst} to solve for the dynamics of the open system (and therefore the purity). 

We follow the formalism of \cite{Colas:2022kfu} and consider the (equal-time) two-point correlation functions of the theory written in terms of the \textit{covariance matrix} $\boldsymbol{\Sigma}^{\mfs}_{\bm{k}}$ with components\footnote{{\it c.f.}~Eq.~(\ref{PowerSpectrum_def}) which shows that $\Sigma^{\mfs}_{\bm{k},11} = 2 \epsilon a^2 M_{\mathrm{Pl}}^2 \cdot \frac{2\pi^2}{k^3} \cdot P_{\zeta}(k)$ and $\Sigma^{\mfs}_{\bm{k},33} = a^2 \cdot \frac{2\pi^2}{k^3} \cdot P_{\sigma}(k)$.}
\begin{eqnarray} \label{corr_Eq_def}
\mathrm{Tr}\big[  \tfrac{1 }{2}  \left\{ z^{\mfs}_{\bm{k},i}(\eta) , z^{\mfs}_{\bm{k},j}(\eta) \right\} \rho^{\mfs}_{\bm{k}}(\eta) \big]  \ = : \ \Sigma^{\mfs}_{\bm{k}, ij}(\eta) \times \frac{\mathcal{V}}{(2\pi)^3} \ .
\end{eqnarray}
Upon differentiating the full Hamiltonian (\ref{H_def}) one finds that the equations of motion for the $4\times 4$ matrix $\boldsymbol{\Sigma}^{\mfs}_{\bm{k}}$ are given by 
\begin{eqnarray} \label{Transport_exact}
\frac{\partial \mathbf{\Sigma}^{\mfs}_{\bm{k}}}{\partial \eta} \ = \ \mathbf{\Omega} \mathbb{H}_{\bm{k}} \mathbf{\Sigma}^{\mfs}_{\bm{k}} - \mathbf{\Sigma}^{\mfs}_{\bm{k}} \mathbb{H}_{\bm{k}} \mathbf{\Omega}  \qquad \mathrm{with} \quad \mathbf{\Omega} = \left[ \begin{smallmatrix} 0 & +1 & 0 & 0 \\ -1 & 0 & 0 & 0 \\ 0 & 0 & 0 & +1 \\ 0 & 0 & -1 & 0 \\ \end{smallmatrix} \right] \,,
\end{eqnarray}
where $\mathbb{H}_{\bm{k}}(\eta)$ is the Hamiltonian matrix from Eq.~(\ref{H_matrix}).  Note that the matrix $\boldsymbol{\Sigma}^{\mfs}_{\bm{k}}$ is symmetric by construction so there are in fact only 10 degrees of freedom in the above. We impose the standard Bunch-Davies initial conditions such that $\Sigma^{\mfs}_{11,\bm{k}}(\eta_{\mathrm{in}}) = \Sigma^{\mfs}_{33,\bm{k}}(\eta_{\mathrm{in}}) = \frac{1}{2k}$ and $\Sigma^{\mfs}_{22,\bm{k}}(\eta_{\mathrm{in}}) = \Sigma^{\mfs}_{44,\bm{k}}(\eta_{\mathrm{in}}) = \frac{k}{2}$ with all other components initially vanishing. 

Since the perturbative theory here is Gaussian (to the level that we are working at), the quantum state of the adiabatic mode is fully described by the system covariance matrix $\Sigma_{\ssA\bm{k}}^{\mfs}$, defined in terms of the correlators of the adiabatic mode,
\begin{eqnarray} \label{system_corr_mat}
\mathbf{\Sigma}_{\ssA\bm{k}}^{\mfs} : = \left[ \begin{matrix} \Sigma^{\mfs}_{11,\bm{k}} & \Sigma^{\mfs}_{12,\bm{k}} \\ \Sigma^{\mfs}_{21,\bm{k}} & \Sigma^{\mfs}_{22,\bm{k}} \end{matrix} \right] \ .
\end{eqnarray}
With this expression in mind, it turns out that the purity takes the simple form \cite{Serafini:2003ke, Martin:2021znx, Colas:2021llj, Martin:2022kph},
\begin{eqnarray} \label{purity_det_rel}
\gamma_{\bm{k}}^{\mfs}  \ = \ \frac{1}{\sqrt{ 4 \det \mathbf{\Sigma}_{\ssA\bm{k}}^{\mfs} }}  \ = \ \frac{1}{2 \sqrt{ \Sigma^{\mfs}_{11,\bm{k}} \Sigma^{\mfs}_{22,\bm{k}} - \Sigma^{\mfs}_{12,\bm{k}} \Sigma^{\mfs}_{21,\bm{k}}  }} \ . 
\end{eqnarray}
In order to numerically evaluate the purity using the above formula, we integrate the transport equations (\ref{Transport_exact}) as well as the equation of motion for $ \det \mathbf{\Sigma}_{\ssA\bm{k}}^{\mfs}$, which in the present instance takes the form (making use of the relation (\ref{Transport_exact}))
\begin{eqnarray} \
\frac{ \partial \det \mathbf{\Sigma}_{\ssA\bm{k}}^{\mfs}}{\partial \eta} & = & \frac{ \partial \Sigma_{11,\bm{k}}^{\mfs}}{\partial \eta} \Sigma^{\mfs}_{22,\bm{k}} + \Sigma^{\mfs}_{11,\bm{k}} \frac{ \partial \Sigma_{22,\bm{k}}^{\mfs}}{\partial \eta}  - 2 \frac{ \partial \Sigma_{12,\bm{k}}^{\mfs}}{\partial \eta} \Sigma_{12,\bm{k}}^{\mfs} \\
& = & 2 \lambda(\eta) \bigg[ \Sigma^{\mfs}_{12,\bm{k}} \Sigma^{\mfs}_{23,\bm{k}} - \Sigma^{\mfs}_{13,\bm{k}} \Sigma^{\mfs}_{22,\bm{k}}  \bigg]\,. \label{det_exact}
\end{eqnarray}
We note that technically (\ref{det_exact}) is a redundant equation, but it is most efficient to numerically integrate the above alongside (\ref{Transport_exact}). The initial condition is $\det\mathbf{\Sigma}_{\ssA\bm{k}}^{\mfs}(\eta_{\mathrm{in}}) = \frac{1}{4}$ which means the purity satisfies $\gamma_{\bm{k}}^{\mfs} (\eta_{\mathrm{in}}) = 1$ as should be the case starting from a pure state.

\subsection{Perturbative treatment}

In this section we assume the interaction $H_{\mathrm{int}}$ defined in (\ref{Hint_def}) is sufficiently mild such that one can apply perturbation theory -- in practice\footnote{Note that our later analysis shows (see the text surrounding Eq.~(\ref{massless_add_constraint}) for example) that the purity in perturbation theory is a good approximation when $|I_{\bm{k}}|^2 \ll 1$ with $I_{\bm{k}}$ defined in (\ref{Ik_def}). This translates to $ \frac{\sqrt{2\epsilon} f_1 }{ (\epsilon - 1) } (- k \eta)^{-N}  \ll 1$ for a power $N$ generally depending on the effective mass of the isocurvature mode and/or $\epsilon$. At early times this is trivially satisfied, however breaks down at late enough times. When decoherence proceeds efficiently our perturbative resummation in (\ref{purity_det_rel_2}) applies, and it turns out that the weaker condition in (\ref{weak_int}) is required to approximate the dynamics. We take this to mean we are in the perturbative regime for the remainder of this work.} we take this to mean that the dimensionless coefficient appearing in the coupling $\lambda(\eta)$ defined in (\ref{coupling_def}) is small such that
\begin{eqnarray} \label{weak_int}
\left| \frac{\sqrt{2\epsilon} f_1 }{ \epsilon - 1 } \right| \ \ll \  1  \ .
\end{eqnarray} 
To begin, we summarize the free evolution determined by the system and environment free Hamiltonians $H_{\ssA}$ and $H_{\ssB}$, defined in (\ref{HA_def}) and (\ref{HB_def}) respectively. The Heisenberg equations of motion yield
\begin{equation} \label{Heis_EoM}
  \begin{split}
    \dfrac{\exd v^{\mfs}_{\bm{k}} }{\exd \eta} &= p^{\mfs}_{\bm{k}} + \frac{a'}{a} v^{\mfs}_{\bm{k}} \\
    \dfrac{\exd p^{\mfs}_{\bm{k}} }{\exd \eta} & = - \frac{a'}{a} p^{\mfs}_{\bm{k}} - k^2 v^{\mfs}_{\bm{k}}
  \end{split}
\qquad\qquad\qquad\qquad
  \begin{split}
    \dfrac{\exd \chi^{\mfs}_{\bm{k}}(\eta) }{\exd \eta} &= \pi^{\mfs}_{\bm{k}} + \frac{a'}{a} \chi^{\mfs}_{\bm{k}} \\
    \dfrac{\exd \pi^{\mfs}_{\bm{k}}(\eta) }{\exd \eta} &= - \frac{a'}{a} \pi^{\mfs}_{\bm{k}} - [ k^2 + M_{\mathrm{eff}}^2(\eta) ] \chi^{\mfs}_{\bm{k}} \,.
  \end{split}
\end{equation}
This means in particular that
\begin{equation}
\dfrac{\exd^2 v^{\mfs}_{\bm{k}} }{\exd \eta^2} + \bigg( k^2 + \frac{\epsilon-2}{(\epsilon - 1)^2 \eta^2} \bigg) v^{\mfs}_{\bm{k}} \ = \ 0\,,
\end{equation}
where we have used the explicit form of the scale factor (\ref{scalefactor}) in the above. Expanding the operator $v^{\mfs}_{\bm{k}}$ in terms of ladder operators in the system sector\footnote{Note the label $\mfs = \mathrm{R},\mathrm{I}$ on the ladder operators, defined as $\mathfrak{a}^{\mathrm{R}}_{\bm{k}} := \frac{\mathfrak{a}_{\bm{k}} + \mathfrak{a}_{-\bm{k}}}{\sqrt{2}}$ and $\mathfrak{a}^{\mathrm{I}}_{\bm{k}} := \frac{\mathfrak{a}_{\bm{k}} - \mathfrak{a}_{-\bm{k}}}{\sqrt{2} i}$ ({\it c.f.}~Eq.~(\ref{RI_label_def})) in terms of the usual ladder operators (satisfying $[\mathfrak{a}_{\bm{k}},\mathfrak{a}^{\dagger}_{\bm{q}}] = \delta(\bm{k} - \bm{q})$) and obey $[\mathfrak{a}^{\mfs}_{\bm{k}},\mathfrak{a}^{\mfr\dagger}_{\bm{q}}] = \delta_{\mfs\mfr}\delta(\bm{k} - \bm{q})$.} 
\begin{eqnarray}
v^{\mfs}_{\bm{k}}(\eta) & = & \mathrm{v}_{\bm{k}}(\eta) \mathfrak{a}^\mfs_{\bm{k}} + \mathrm{v}^{\ast}_{\bm{k}}(\eta) \mathfrak{a}_{\bm{k}}^{\mfs\dagger}\,,
\end{eqnarray}
implies that the mode functions have the form
\begin{eqnarray} \label{vmodes_def}
\mathrm{v}_{\bm{k}}(\eta) \ = \ \sqrt{ \frac{\pi}{4k} } \; e^{+ i \tfrac{\pi}{2}\gamma + i \tfrac{\pi}{4}}  \; \sqrt{ - k \eta } H^{(1)}_{\gamma}(-k \eta) \qquad \mathrm{with}\ \gamma := \frac{\epsilon - 3}{2 ( \epsilon - 1 ) }\,,
\end{eqnarray}
where $H_{\gamma}^{(1)}$ is the Hankel function of the first kind. This is a generalization of the inflationary result which is indeed recovered in the limit $\epsilon \ll 1$. Note that the above uses the Bunch-Davies initial conditions where $\mathrm{v}_{\bm{k}}(\eta) \simeq \frac{e^{- i k \eta}}{\sqrt{2k}}$ for $-k \eta \gg 1$. 

Since the interaction Hamiltonian (\ref{Hint_def}) consists of interactions $p^{\mfs}_{\bm{k}}(\eta) \otimes \chi^{\mfs}_{\bm{k}}(\eta)$, we consider the expansion of these two field operators in terms of ladder operators of the system sector and environment sector:
\begin{eqnarray}
p^{\mfs}_{\bm{k}}(\eta) & = & \mathrm{p}_{\bm{k}}(\eta) \mathfrak{a}^\mfs_{\bm{k}} + \mathrm{p}^{\ast}_{\bm{k}}(\eta) \mathfrak{a}_{\bm{k}}^{\mfs\dagger} \\
\chi^{\mfs}_{\bm{k}}(\eta) & = & \mathrm{x}_{\bm{k}}(\eta) \mathfrak{b}^{\mfs}_{\bm{k}} + \mathrm{x}^{\ast}_{\bm{k}}(\eta) \mathfrak{b}_{\bm{k}}^{\mfs\dagger}\,.
\end{eqnarray}
This implies that the Bunch-Davies state (\ref{BD}) should be decomposed as 
\begin{eqnarray} \label{BD_dec}
 |\mathrm{BD} \rangle : = | \mathrm{BD}_{\ssA} \rangle \otimes  |\mathrm{BD}_{\ssB} \rangle\,,
\end{eqnarray}
such that $\mathfrak{a}_{\bm{k}}^{\mfs}|\mathrm{BD}_{\ssA} \rangle = 0$ and $\mathfrak{b}_{\bm{k}}^{\mfs}|\mathrm{BD}_{\ssB} \rangle = 0$ -- this observation makes it clear that the two sectors begin in a (pure) initial state which is uncorrelated.

Useful for later computations, the Heisenberg equations of motion then tell us that
\begin{eqnarray} \label{pmodes_def}
\mathrm{p}_{\bm{k}}(\eta) \ = \ - \; \sqrt{ \frac{\pi k}{4} }  e^{+ i \tfrac{\pi}{2}\gamma + i \tfrac{\pi}{4}} \sqrt{ - k \eta } H^{(1)}_{\gamma-1}(- k \eta)\,,
\end{eqnarray}
with $\gamma$ defined in (\ref{vmodes_def}) above, as well as\footnote{This stems from the equation of motion $\partial_\eta^2 \chi^{\mfs}_{\bm{k}}  + \left( k^2 + \frac{\epsilon-2 + 12h_2 - 4 \epsilon (f_2+h_2)}{(\epsilon - 1)^2 \eta^2} \right) \chi^{\mfs}_{\bm{k}} = 0$.}
\begin{eqnarray} \label{xmodes_def}
\mathrm{x}_{\bm{k}}(\eta) = \sqrt{ \frac{\pi}{4k} } e^{-  \tfrac{\pi}{2}\mu + i \tfrac{\pi}{4}}  \; \sqrt{ - k \eta } H^{(1)}_{i \mu}(- k \eta) \qquad \mathrm{with} \quad \mu := \sqrt{ \frac{24 h_2 - 8 \epsilon(f_2+h_2) - (\epsilon - 3)^2}{4(\epsilon-1)^2} } \; . \quad
\end{eqnarray}
Note that the index $\mu$ can take on real or pure imaginary values in general.

\subsubsection{Dyson Series and the perturbative purity}

Iteratively solving the von Neumann equation (\ref{vN_EQ}) for the full density matrix gives the standard time-ordered Dyson series:
\begin{equation} \label{Dyson}
\rho(\eta) \simeq \rho(\eta_{\mathrm{in}}) - i \int_{\eta_{\mathrm{in}}}^{\eta} \exd \eta_{1} \left[ H_{\mathrm{int}}(\eta_1), \rho(\eta_{\mathrm{in}}) \right] - \int_{\eta_{\mathrm{in}}}^{\eta} \exd \eta_{1} \int_{\eta_{\mathrm{in}}}^{\eta_1} \exd \eta_{2}\; \left[ H_{\mathrm{int}}(\eta_2), \big[  H_{\mathrm{int}}(\eta_1), \rho(\eta_{\mathrm{in}}) \big] \right] + \ldots
\end{equation}
Using the decomposition (\ref{rho_decomp}) of the density matrix, the Dyson series on Eq.~(\ref{vN_mode}) and performing a trace over the environment we obtain (H.c. denotes Hermitian conjugate)
\begin{eqnarray} \label{PT_rhoA}
 \varrho_{\ssA\bm{k}}^{\mfs}(\eta) & \simeq &  |\mathrm{BD}_{\ssA}\rangle \langle \mathrm{BD}_{\ssA} | \\
 & \ & + \frac{(2\pi)^3}{\mathcal{V}} \int_{\eta_{\mathrm{in}}}^{\eta} \exd \eta_1 \int_{\eta_{\mathrm{in}}}^{\eta_1} \exd \eta_2 \; \Scale[0.89]{ \lambda(\eta_1) \lambda(\eta_2) \bigg( \mathcal{W}^{\mfs}_{\ssB\bm{k}}(\eta_1,\eta_{2}) \big[  p^{\mfs}_{\bm{k}}(\eta_2)   |\mathrm{BD}_{\ssA}\rangle \langle \mathrm{BD}_{\ssA} | , p^{\mfs}_{\bm{k}}(\eta_1) \big] + \mathrm{H.c.} \bigg) } + \mathcal{O}(\lambda^4) \,, \notag
\end{eqnarray}
with environment correlator defined by
\begin{eqnarray}
\mathrm{Tr}_{\ssB}\left[\; \Scale[0.85]{ |\mathrm{BD}_{\ssB}\rangle \langle \mathrm{BD}_{\ssB} | }\; \chi_{\bm{k}}^{\mfs'}(\eta_1) \chi_{\bm{k}'}^{\mfs}(\eta_2)  \right] & = : & \delta_{\mfs\mfs'} \delta^{(3)}(\bm{k} - \bm{k}') \; \mathcal{W}^{\mfs}_{\ssB\bm{k}}(\eta_1,\eta_{2}) \,,
\end{eqnarray}
using the Bunch-Davies initial condition (\ref{BD_dec}), and truncating at second order in the coupling\footnote{Notice that the environmental mass shift proportional to $\lambda^2 \chi^2$ in (\ref{Hint_def}) does not contribute to the dynamics of the open system.}. Note that the formula (\ref{PT_rhoA}) uses the symmetry $\mathcal{W}^{\mfs}_{\ssB\bm{k}}(\eta_2,\eta_1) = \mathcal{W}^{\mfs\ast}_{\ssB\bm{k}}(\eta_1,\eta_2)$ which follows from Hermicity of the field operators. 

Squaring and tracing the above operator, it is then straightforward to derive a time-ordered expression for the purity (\ref{purity_def}) in perturbation theory where
\begin{eqnarray} \label{PP_ordered_pre}
\gamma_{\bm{k}}^{\mfs}(\eta)  & = & 1 - 4 \int_{\eta_{\mathrm{in}}}^{\eta} \exd \eta_1 \int_{\eta_{\mathrm{in}}}^{\eta_1}  \exd \eta_{2}\  \lambda(\eta_1) \lambda(\eta_2)\;  \mathrm{Re}\left[ \mathcal{W}^{\mfs}_{\ssA\bm{k}}(\eta_1,\eta_2)  \mathcal{W}^{\mfs}_{\ssB\bm{k}}(\eta_1,\eta_2) \right]\,,
\end{eqnarray}
where we denote the system correlator
\begin{eqnarray}
\mathrm{Tr}\left[ \Scale[0.85]{ |\mathrm{BD}_{\ssA}\rangle \langle \mathrm{BD}_{\ssA} | }\; p_{\bm{k}}^{\mfs}(\eta_1) p_{\bm{k}'}^{\mfs'}(\eta_2) \right] & = : & \delta_{\mfs\mfs'} \delta^{3}(\bm{k} - \bm{k}') \; \mathcal{W}^{\mfs}_{\ssA\bm{k}}(\eta_1,\eta_{2}) \ .
\end{eqnarray}
A more useful formula writes (\ref{PP_ordered_pre}) as an {\it unnested} double-integral (see also Eq.~(\ref{det_Pert_App})) where\footnote{Notice that $\Sigma^{\mfs}_{\bm{k},22}(\eta) \simeq \mathcal{W}^{\mfs}_{\ssA\bm{k}}(\eta,\eta)$ and $\Sigma^{\mfs}_{\bm{k},33}(\eta) \simeq \mathcal{W}^{\mfs}_{\ssB\bm{k}}(\eta,\eta)$ at early times in perturbation theory.}
\begin{eqnarray} \label{PP_ordered}
\gamma_{\bm{k}}^{\mfs}(\eta) & = & 1 - 2 \int_{\eta_{\mathrm{in}}}^{\eta} \exd \eta_1  \int_{\eta_{\mathrm{in}}}^{\eta} \exd \eta_{2}\  \lambda(\eta_1) \lambda(\eta_2) \mathcal{W}^{\mfs}_{\ssA\bm{k}}(\eta_1,\eta_2)  \mathcal{W}^{\mfs}_{\ssB\bm{k}}(\eta_1,\eta_2)  \ .
\end{eqnarray}
In the specific case of interest here, the correlators are easily expressed in terms of the mode functions $\mathrm{p}_{\bm{k}}(\eta)$ in (\ref{pmodes_def}) and $\mathrm{x}_{\bm{k}}(\eta)$ in (\ref{xmodes_def}) such that
\begin{eqnarray} \label{W_modes}
\mathcal{W}^{\mfs}_{\ssA\bm{k}}(\eta_1,\eta_2) = \mathrm{p}_{\bm{k}}(\eta_1)  \mathrm{p}^{\ast}_{\bm{k}}(\eta_2) \qquad \mathrm{and} \qquad \mathcal{W}^{\mfs}_{\ssB\bm{k}}(\eta_1,\eta_2) = \mathrm{x}_{\bm{k}}(\eta_1)  \mathrm{x}^{\ast}_{\bm{k}}(\eta_2)\,.
\end{eqnarray}
This leads to the following perturbative expression for the purity,
\begin{eqnarray} \label{PP_ordered_2}
\gamma_{\bm{k}}^{\mfs}(\eta)  & = & 1 - 2 \int_{\eta_{\mathrm{in}}}^{\eta} \exd \eta_1  \int_{\eta_{\mathrm{in}}}^{\eta} \exd \eta_{2}\  \lambda(\eta_1) \lambda(\eta_2) \mathrm{p}_{\bm{k}}(\eta_1)\mathrm{p}^{\ast}_{\bm{k}}(\eta_2) \mathrm{x}_{\bm{k}}(\eta_1)\mathrm{x}^{\ast}_{\bm{k}}(\eta_2) \ = \ 1 - 2 \big| I_{\bm{k}}(\eta) \big|^2 \,,\qquad
\end{eqnarray}
with the definition
\begin{equation} \label{Ik_def}
I_{\bm{k}}(\eta) \ : = \ \int_{\eta_{\mathrm{in}}}^{\eta} \exd \eta' \; \lambda(\eta') \mathrm{p}_{\bm{k}}(\eta') \mathrm{x}_{\bm{k}}(\eta')  \ .
\end{equation}

\subsubsection{Late-time perturbative breakdowns and a simple resummation}

We note here that time-dependent perturbation theory generically fails at late-times. Fundamentally, this is because expressions such as (\ref{Dyson}) above are the result of expanding unitary time-evolution operators (schematically) as $e^{- i H_{\mathrm{int}} t} \simeq 1 - i H_{\mathrm{int}} t + \ldots$ -- these are series expansions whose corrections grow without bound for times $H_{\mathrm{int}} t \sim \mathcal{O}(1)$ or larger. This issue of {\it secular growth} is a generic problem that especially plagues quantum mechanics in curved and/or time-dependent backgrounds, which we find also occurs in calculations of the perturbative purity in this work.

This doesn't necessarily pose an obstacle to making reliable late-time predictions: in many cases where secular growth occurs, one may perform a late-time resummation. This typically entails extending the domain of validity of perturbative expressions (such as (\ref{Dyson})) to integro-differential equations governing the reduced density matrix, often referred to as master equations \cite{Lindblad:1975ef, breuerTheoryOpenQuantum2002, breuerTimelocalMasterEquations2002, whitneyStayingPositiveGoing2008, Scarlatella:2021vfr, Burgess:2022rdo, Colas:2022hlq} -- these operate in a manner analogous to renormalization group flows, whose scale-dependence is first derived in perturbation theory and then extended to larger domains by differentiation with respect to scale (or time in this instance). 

Since the goal of this work is to determine when decoherence occurs, it turns out that master equations are overkill for the examples of secular growth encountered in this work. A simple resummation is obtained by noting that the determinant of the system covariance matrix takes the form
\begin{eqnarray}
 \det \boldsymbol{\Sigma}^{\mfs}_{\ssA\bm{k}}  \; \big|_{\mathrm{PT}} & \simeq & \frac{1}{4} + \big| I_{\bm{k}}(\eta) \big|^2 \,,
\end{eqnarray}
in perturbation theory (for a derivation see Appendix \ref{App:det_PT}). We note that the relationship (\ref{purity_det_rel}) is true to all orders in the coupling, and so an excellent resummation for the purity dynamics is given by 
\begin{eqnarray} \label{purity_det_rel_2}
\gamma_{\bm{k}}^{\mfs}(\eta)  \ = \ \frac{1}{\sqrt{ 4 \det \mathbf{\Sigma}_{\ssA\bm{k}}^{\mfs}(\eta) }}  \ \simeq \ \frac{1}{\sqrt{ 1 + 4 | I_{\bm{k}}(\eta) |^2 } } \,,
\end{eqnarray}
which agrees with (\ref{PP_ordered_2}) to sub-leading order when $I_{\bm{k}}(\eta)$ is small (as we see below in {\it e.g.}~Figure \ref{figure:Fig3}). Since $\gamma_{\bm{k}}^{\mfs} \propto ( \det \mathbf{\Sigma}_{\ssA\bm{k}}^{\mfs})^{-1/2}$ is true to all orders in perturbation theory, this allows (\ref{purity_det_rel_2}) to resum perturbation theory to all orders in $| I_{\bm{k}}(\eta) |^2 \propto \lambda^2(\eta) $.

\section{Purity freezing \textit{vs.}~decoherence in inflation}
\label{sec:inflation}

In this section we recap some results of Ref.~\cite{Colas:2022kfu}, which studies the model at hand in the case of an expanding phase with $\epsilon \ll 1$ (corresponding to the linear regime of quasi-single field inflation). After investigating features of the purity in the exact description ({\it i.e.}~numerically), we consider weak couplings $f_1$ and compare exact descriptions to expressions in perturbation theory. Resumming perturbation theory better approximates the late-time evolution (although as we will see fails to capture all the features of the exact description) and is analytically powerful. Using the resummation we are able to analytically determine when decoherence proceeds (for light environments generically), as well as derive a novel formula for the value of purity freezing in the heavy environment limit. 

\subsection{Exact purity}

In what follows, we start by providing some results of numerically integrating the exact dynamics, and based off of this evidence briefly summarize when decoherence does and does not occur. 

Recall from Eq.~(\ref{full_tdepmass}) that the exact mass of the environment becomes shifted through the system-environment interaction to $M_{\mathrm{eff}}^2(\eta) + \lambda^2(\eta)$ with $M_{\mathrm{eff}}^2(\eta)$ defined in (\ref{eff_mass_sigma}) and the coupling defined in (\ref{coupling_def}). In the inflationary case we may interpret the effective mass as related to the ratio of an ordinary mass parameter $m$ to the Hubble scale $H$, {\it i.e.}
\begin{eqnarray}
M_{\mathrm{eff}}^2(\eta) \ = \ \frac{m^2}{H^2 \eta^2} \qquad \mathrm{with}\qquad \frac{m^2}{H^2} :=  6 h_2  \,, 
\end{eqnarray}
where we have also used $\epsilon \ll 1$. This interpretation will be useful to compare with \cite{Colas:2022kfu}.

\begin{figure}[h]
\centering
  \includegraphics[width=0.70\linewidth]{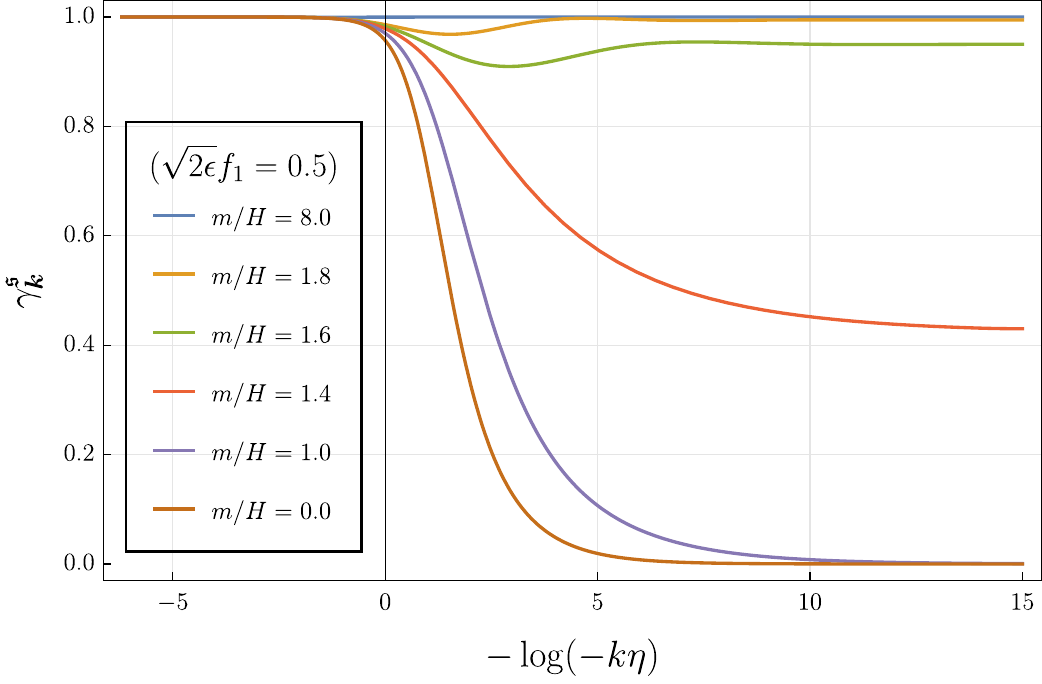}
\caption{Exact evolution for the purity $\gamma_{\bm{k}}^\mfs$ as a function of $-\log(-k\eta)$ (in the inflationary $\epsilon \ll 1$ case) for various values of $m/H$. The above assumes coupling $\sqrt{2\epsilon} f_1 = 0.5$.}
\label{figure:Fig1}
\end{figure}

We here integrate the exact transport equations (\ref{corr_Eq_def}) alongside the (redundant) equation of motion for $\det\boldsymbol{\Sigma}_{\ssA\bm{k}}^{\mfs}$ in (\ref{det_exact}) for various values of $m/H$. The exact evolution of the purity is shown in Figure~\ref{figure:Fig1}, for various values of $m/H$, integrated in terms of cosmic time $t$ defined through the relation\footnote{The definition (\ref{scalefactor}) of the scale factor in the present instance obviously implies that $\eta_0 = - 1/H$ in the case of inflation where $\epsilon \ll 1$.}
\begin{eqnarray}
\exd t = a \exd \eta \qquad \implies \qquad \eta(t) = - \frac{1}{H} e^{-Ht} \ .
\end{eqnarray}
Interestingly, Figure~\ref{figure:Fig1} shows that the adiabatic mode undergoes decoherence (with purity approaching zero at late times) only for light environment fields satisfying roughly $m/H \lesssim \mathcal{O}(1)$ (the precise bound being $0 \leq m \leq \sqrt{2}H$, as computed in Eq.~(\ref{eq:bound})  below). As one increases the mass of the isocurvature mode, one finds that the purity freezes to a non-zero value (somewhere between $0$ and $1$) and there is no late-time decoherence. For relatively heavier masses, the phenomenon of ``recoherence'' is observed where the purity first decreases and then returns close to the initial value 1. As one makes $m/H$ very large, the purity remains infinitesimally close to $1$ -- as described in \cite{Colas:2022kfu}, this is consistent with the decoupling limit of the environment field where one must recover a unitary Wilsonian EFT (where the purity remains 1 for all times). This picture of decoupling is consistent with the finding that for intermediate values of masses ($m \gtrsim \sqrt{2}H$) the purity freezes to some value between 0 and 1.

\begin{figure}[h]
\centering
  \includegraphics[width=0.47\linewidth]{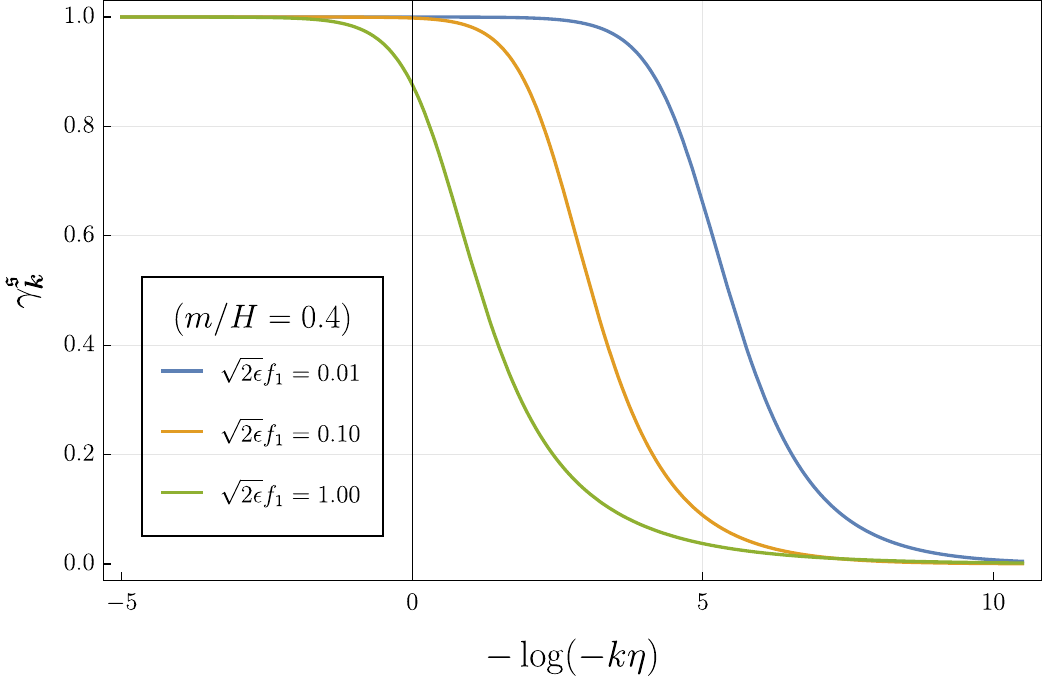} \ \includegraphics[width=0.476\linewidth]{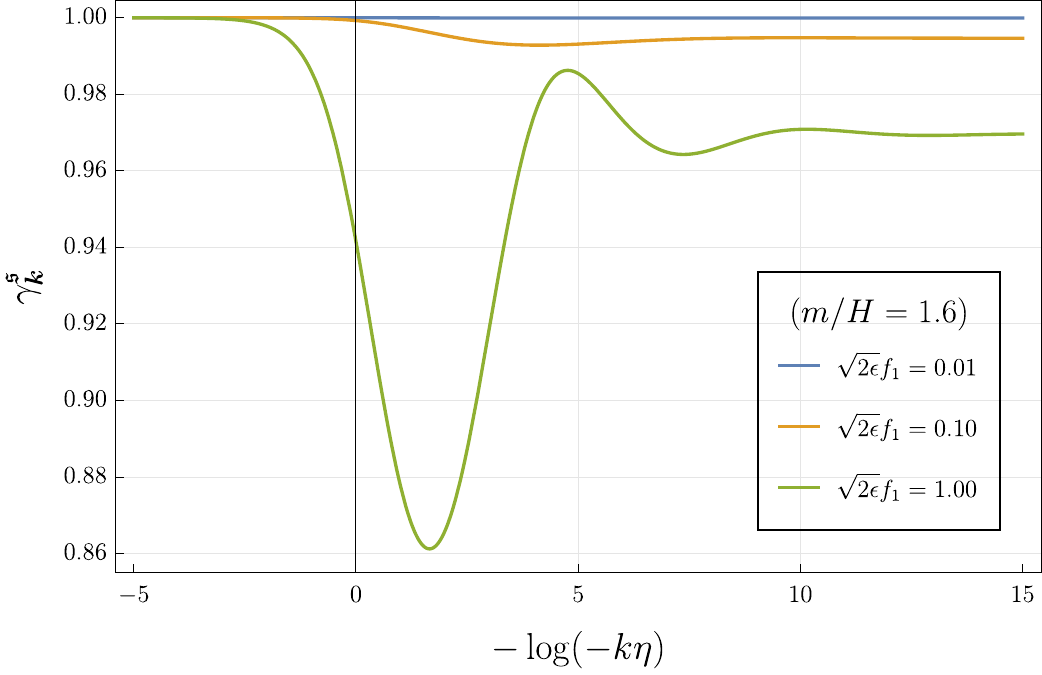}
\caption{Exact evolution for the purity $\gamma_{\bm{k}}^\mfs$ as a function of $-\log(-k\eta)$ (in the inflationary $\epsilon \ll 1$ case) for various values of the coupling $\sqrt{\epsilon} f_1$. Left Panel: $m/H=0.4$ leading to decoherence. Right Panel: $m/H = 1.6$ leading to non-trivial purity freezing at late times.}
\label{figure:Fig2}
\end{figure}

We also show the effect of varying the size of the coupling here (controlled by the size of $\sqrt{\epsilon}f_1$ in the present instance). As Figure~\ref{figure:Fig2} shows, for intermediate values of masses, smaller couplings means the purity tends to be less and less perturbed from the initial value 1 (tending towards the unitary limit where the purity is 1 for all times in the limit of vanishing coupling).

\subsection{Perturbative and resummed purity}

The previous subsection demonstrates the exact behaviour of the purity. We here compare this to the perturbative evolution of the purity in (\ref{PP_ordered_2}), as well as its resummed formulation  where one is able to retain analytical control (unlike in the exact case). 

In the inflationary case with $\epsilon \ll 1$ the system mode functions (\ref{pmodes_def}) simplify to
\begin{eqnarray}
\mathrm{p}_{\bm{k}}(\eta) \ \simeq \ - i \sqrt{ \frac{k}{2} } \; e^{ - i k \eta}\,,
\end{eqnarray}
meanwhile the form of the environment modes functions (\ref{xmodes_def}) remain unsimplified, with the parameter $\mu$ there taking the form\footnote{We here consider $\mu = i \nu$ (where $\nu = \sqrt{  \frac{9}{4} - \frac{m^2}{H^2}} $ is the normally defined order parameter of the Hankel function) as our primary interest is on heavy fields for which $m/H>3/2$ (in the principle series) so $\mu \in \mathbb{R}$. Then, for fields in the complementary series ($0 < m/H < 3/2$), $\mu$ is indeed pure imaginary.}
\begin{eqnarray}
\mu  \ \simeq  \ \sqrt{ 6 h_2 - \frac{9}{4} } \  = \  \sqrt{ \frac{m^2}{H^2} - \frac{9}{4} } \ . 
\end{eqnarray}
The analysis of \cite{Tolley:2007nq} suggests that imposing scale-invariance in case (i) puts no constraints on the parameter $h_2$, which means $\mu$ can be either real or pure imaginary. 

Both the purity in perturbation theory in Eq.~(\ref{PP_ordered_2}) as well as its resummed version in Eq.~(\ref{purity_det_rel_2}) are controlled by the integral $I_{\bm{k}}(\eta)$ defined in (\ref{Ik_def}). The computation of this integral is performed using the above mode functions and coupling (\ref{coupling_def}) in Appendix \ref{App:Ik_inflation}, where one eventually finds (for $\mu \neq \frac{3i}{2}, \frac{i}{2}$ {\it i.e.}~the massless and conformally coupled cases respectively for which the result can also be computed),
\begin{eqnarray} \label{Iksq_body}
\left| I_{\bm{k}}(\eta) \right|^2 =  \epsilon f_1^2 \; \bigg| \; \frac{\pi \; \mathrm{sech}(\pi\mu) }{\sqrt{2}} - e^{- i \tfrac{\pi}{4}} \sqrt{ \frac{ - k \eta}{\pi}  } \; \big[ \; g_{\mu}(- k \eta) + g_{-\mu}(- k\eta) \; \big] \; \bigg|^2 \,,
\end{eqnarray}
with the definition
\begin{equation}
g_{\mu}(z) \ := \ \frac{ \Gamma(- i \mu) e^{ + \tfrac{\pi\mu}{2} }  }{1 + 2 i \mu}\; \left( \frac{z}{2} \right)^{i\mu} \ _2F_2 \;^{\frac{1}{2}+i\mu, \frac{1}{2}+i\mu}_{\frac{3}{2}+i\mu, 1 + 2 i \mu}(2 i z)\,,
\end{equation}
where $_2F_{2}$ is a generalized hypergeometric function. One generically finds at early times that 
\begin{eqnarray}
| I_{\bm{k}}(\eta) |^2 \  \simeq \ \epsilon f_1^2\bigg[ \frac{1}{8 (- k \eta)^2} + \mathcal{O}\left( (- k \eta)^{-4} \right) \bigg] \qquad \qquad \mathrm{for}\  - k \eta \gg 1 \ ,
\end{eqnarray}
and since we have $\gamma_{\bm{k}}^{\mfs}(\eta) \simeq 1 - 2  | I_{\bm{k}}(\eta) |^2$ from Eq.~(\ref{PP_ordered_2}), we can show analytically that the purity starts off at 1 at early times and slowly decreases from there on.
\begin{figure}[h]
\centering
  \includegraphics[width=0.96\linewidth]{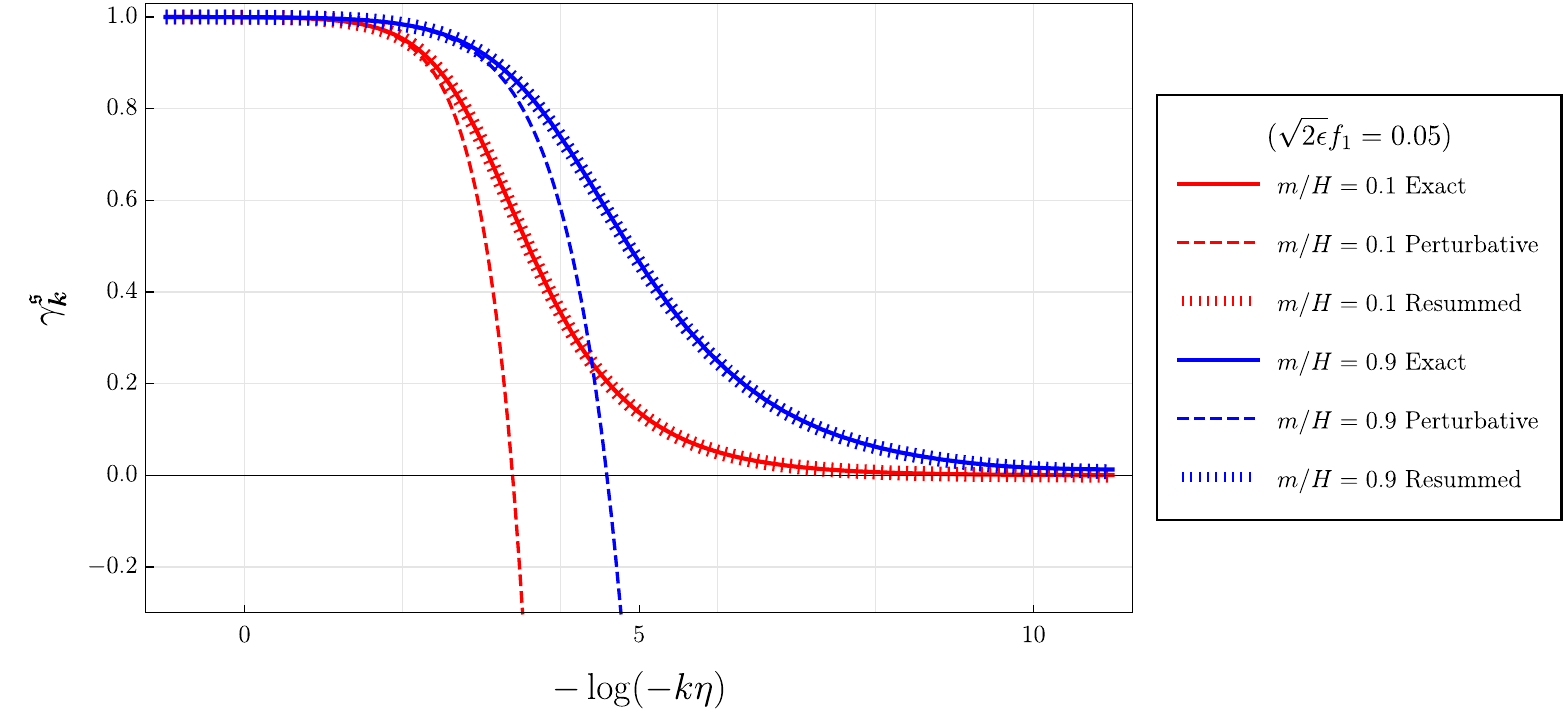}
\caption{Exact vs.~perturbative vs.~resummed evolution for the purity $\gamma_{\bm{k}}^\mfs$ as a function of $-\log(-k\eta)$ (in the inflationary $\epsilon \ll 1$ case) for coupling $\sqrt{2 \epsilon} f_1 = 0.05$ and two choices of $m/H$.}
\label{figure:Fig3}
\end{figure}

In Figure~\ref{figure:Fig3}, we see that for choices of light masses (where decoherence {\it does} occur), the perturbative purity clearly suffers secular growth breakdown as the purity becomes negative at relatively early times -- this occurs since at late times $I_{\bm{k}}(\eta)$ grows without bound and the condition $|I_{\bm{k}}(\eta)| \ll 1$ for perturbativity breaks down. We also see however that the resummation from (\ref{purity_det_rel_2}) performs remarkably well at late times (for comments on the size of the error in the resummation see Appendix \ref{App:MEcomparison}).

This suggests that the resummation (\ref{purity_det_rel_2}) provides a robust analytic expression for the scaling of the purity at late times as long as the coupling is not too large. This is most easily seen in the simplest case of a massless environment with $m/H =0$ (see Eq.~(\ref{Ik_massless})) where
\begin{eqnarray}
\gamma_{\bm{k}}^{\mfs}(\eta) \ \simeq \ \frac{1}{\sqrt{1 + 4 |I_{\bm{k}}(\eta)|^2}} \ = \ \frac{1}{ \sqrt{ 1 + 2 \epsilon f_1^2 \left| \frac{e^{- 2 i k \eta}}{- k \eta} - \pi - i \mathrm{Ei}(-2 i k\eta) \right|^2  } } \qquad  (\mathrm{massless})\,,
\end{eqnarray}
with Ei is the exponential integral function. At late times, the scaling of the function under the square root is given by
\begin{eqnarray} \label{massless_scaling}
 4 |I_{\bm{k}}(\eta)|^2 \ \simeq \ 2 \epsilon f_1^2 \bigg[ \frac{1}{(- k \eta)^2} - \frac{\pi}{(- k \eta)} + \ldots \bigg] \qquad \mathrm{when}\ - k \eta \ll 1 \ .
\end{eqnarray}
What this expression implies is that the late-time scaling of the purity is given by
\begin{eqnarray} \label{massless_gamma_inf}
\gamma_{\bm{k}}^{\mfs}(\eta) \ \simeq \ \frac{1}{\sqrt{2\epsilon f_1}} \bigg[ - k \eta + \mathcal{O}\big( \Scale[0.90]{(-k\eta)^2} \big) \bigg] \qquad \qquad \mathrm{for} \; - k \eta \ll 1 \qquad (\mathrm{massless}) \,,
\end{eqnarray}
given the additional constraint that $ 4 |I_{\bm{k}}(\eta)|^2 \gg 1$, which given (\ref{massless_scaling}) means 
\begin{eqnarray} \label{massless_add_constraint}
 \frac{2 \epsilon f_1^2}{(- k \eta)^2} \gg 1 \ .
\end{eqnarray}
Assuming Eq.~(\ref{weak_int}) one has $\sqrt{2 \epsilon} f_1 \ll 1$ and so the above condition is trivially satisfied in this case ({\it i.e.}~at late enough times one gets the required hierarchy $(- k \eta)^2 \ll 2 \epsilon f_1^2 \ll 1$ to get decoherence). We conclude that a massless environment fully decoheres the system in the perturbative regime at a rate which is exponentially fast in terms of the cosmic time. For a mode spending around $50$ efolds above the horizon, the final value of the purity would be $\gamma_{\bm{k}}^{\mfs}(\eta_{\mathrm{f}}) \sim \ee^{-50}$ which corresponds to a maximally mixed state. 

More generally, we notice that for light environments (in a sense we make precise below), we have full decoherence of the reduced density matrix at late-time as well -- this can be understood analytically by considering the late-time limit $- k \eta \ll 1$ of (\ref{Iksq_body}) for generic $\mu \neq \frac{3i}{2}, \frac{i}{2}$ (or $m/H \neq 0, \sqrt{2}$):
\begin{eqnarray} \label{inflation_Iksq_LT}
&& |I_{\bm{k}}(\eta)|^2 \simeq   \frac{ \epsilon f_1^2 }{2} \; \bigg(  \frac{\Gamma (-i \mu )^2}{\pi  \left(\tfrac{1}{2}+i \mu \right)^2} \left(\frac{-k\eta}{2}\right)^{1+2 i \mu }+\frac{\Gamma (i \mu )^2 }{\pi  \left(\tfrac{1}{2}-i \mu \right)^2} \left(\frac{-k\eta}{2}\right)^{1-2 i \mu } \\
& \ & \quad - \frac{ \sqrt{\pi } \; \Gamma (-i \mu ) }{(\tfrac{1}{2}+i \mu) \sin\big[ \tfrac{\pi}{2} ( \tfrac{1}{2} + i \mu ) \big] } \left(\frac{-k\eta}{2}\right)^{\frac{1}{2}+i \mu }  \; - \; \frac{ \sqrt{\pi } \; \Gamma (i \mu )  }{( \tfrac{1}{2}- i \mu ) \sin\big[ \tfrac{\pi}{2} ( \tfrac{1}{2} - i \mu  ) \big]} \left(\frac{-k\eta}{2}\right)^{\frac{1}{2}-i \mu } + \pi^2 \mathrm{sech}^2(\pi\mu) \bigg)\,. \notag
\end{eqnarray}
Which term dominates in the above expression depends on the size of $m/H$. To determine for which range of masses decoherence occurs, it is sufficient to focus on the first term (controlled by the power $1 + 2 i \mu$) which implies that late-time decoherence occurs for
\begin{eqnarray}\label{eq:bound}
0 <m^2< 2 H^2\; ,
\end{eqnarray}
ensuring that $|I_{\bm{k}}(\eta)|^2 $ increases at late enough times. Note that the fastest late-time scaling implied by the above is $|I_{\bm{k}}(\eta)|^2 \propto (-k\eta)^{-2}$ for $m/H \ll 1$, which gives the result (\ref{massless_gamma_inf}). The slowest late-time scaling is obtained for $m \rightarrow \sqrt{2}H$ with $|I_{\bm{k}}(\eta)|^2 \propto \log^2 (-k\eta)$, which gives $\gamma_{\bm{k}}^{\mfs}(\eta_{\mathrm{f}}) \sim 1/50$ for 50 efolds above the horizon\footnote{This result can easily be obtained by replacing the massive mode functions Eq.~(\ref{xmodes_def}) with the ones of a conformally coupled scalar $\mathrm{x}_{\bm{k}}(\eta) \ = \ \frac{1}{\sqrt{2k}} \ee^{-ik\eta}$. See the $- k \eta \ll 1$ expansion of Eq.~(\ref{eq:conformal}).}.  Notice that in order for $|I_{\bm{k}}(\eta)|^2 \gg 1$ to hold in this case (so that the state decoheres) one requires an additional constraint of the form ({\it c.f.}~Eq.~(\ref{massless_add_constraint}))
\begin{eqnarray}
2\epsilon f_1^2 \cdot \frac{\Gamma (-i \mu )^2}{\pi \left(\tfrac{1}{2}+i \mu \right)^2} \left(\frac{-k\eta}{2}\right)^{1+2 i \mu } \gg 1\,.
\end{eqnarray}
Note that this constraint is trivially satisfied when the power of $- k \eta$ is between $-2$ and $-1$ -- once this power gets close to zero (for masses close to $m^2 \rightarrow 2 H^2$), one must additionally enforce this constraint otherwise the resummation may be ruined ({\it i.e.}~if the coupling is too small, the growth of $ |I_{\bm{k}}(\eta)|^2$ can be so slow that $\mathcal{O}(f_1^4)$ effects may become important before it dominates). The conclusion is then that decoherence proceeds efficiently for light masses $m/H \lesssim \mathcal{O}(1)$.\\

\begin{figure}[h]
\centering
\includegraphics[width=0.70\linewidth]{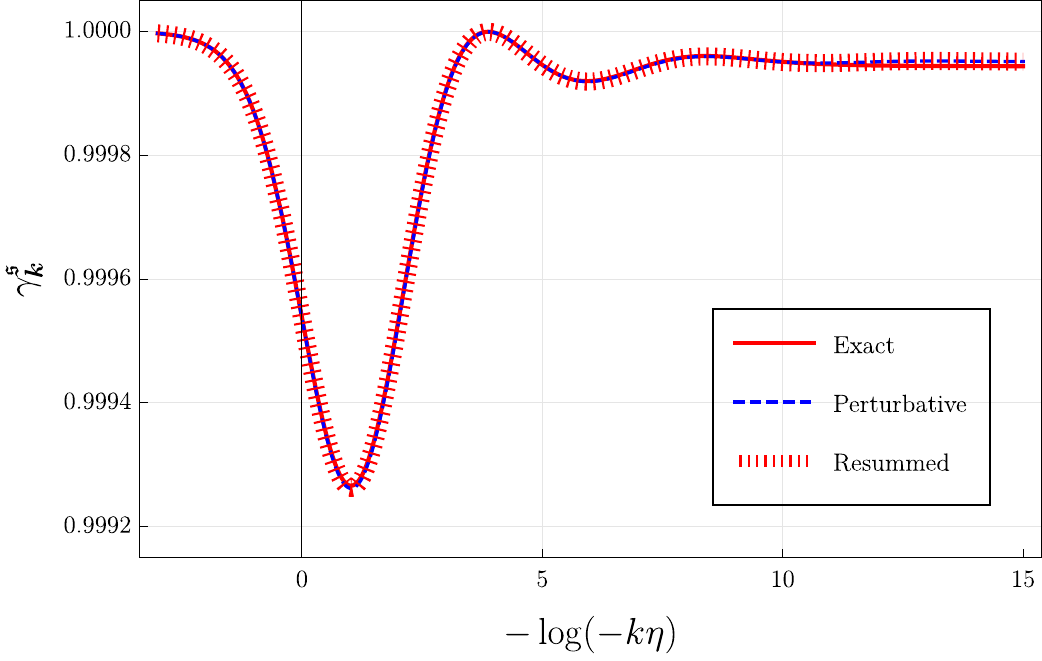}
\caption{Exact vs.~perturbative vs.~resummed evolution for the purity $\gamma_{\bm{k}}^\mfs$ as a function of $-\log(-k\eta)$ (in the inflationary $\epsilon \ll 1$ case) for the choice $m^2 = 4H^2$ and $\sqrt{2 \epsilon} f_1 = 0.1$.}
\label{figure:Fig5}
\end{figure}

\begin{figure}[h]
\centering
\includegraphics[width=0.50\linewidth]{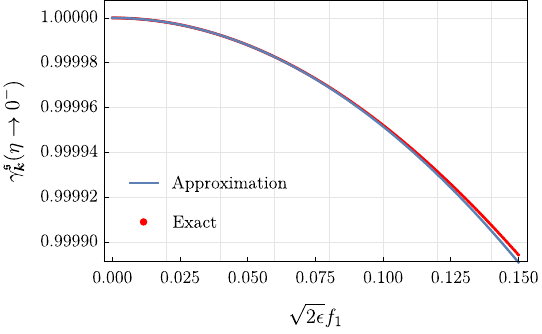} \includegraphics[width=0.49\linewidth]{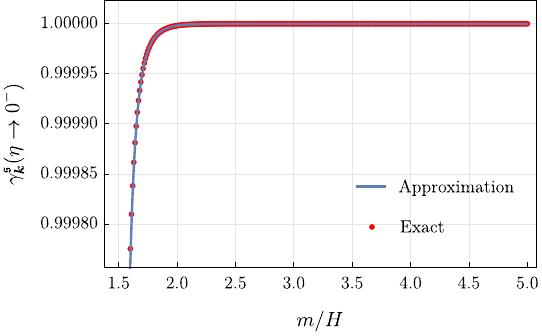}
\caption{Final purity value $\gamma_{\bm{k}}^{\mfs}(\eta \to 0^{-})$ from exact integration tested against the approximation in Eq.~(\ref{heavy_freeze}). Left Panel: as a function of the coupling $\sqrt{2\epsilon}f_1$ (for fixed $m/H = 2$). Right Panel: as a function of $m/H$ (for fixed coupling  $\sqrt{2\epsilon}f_1 =0.02$). We test the expression only until $m/H \sim 5$ as numerical noise due to precision loss begins to dominate for large masses.}
\label{fig:freeze}
\end{figure}

In the opposing limit of very large masses $m/H \gg 1$, the resummation {\it and} perturbation theory become excellent approximations to exact dynamics. This is demonstrated in Figure~\ref{figure:Fig5}, where we can already see that the choice $m^2 = 4 H^2$ represents a sufficiently large mass, with agreement only getting better as the mass is dialled to heavier values. The reason for this is that letting the mass get infinitely large means that we approach the limit where the environment field decouples, and so the purity gets arbitrarily close to 1.
The heavy mass regime therefore bodes well for analytic control of the late-time regime as well: in the case where $m/H \gg 1$ the late-time asymptote of $I_{\bm{k}}(\eta)$ provides an excellent approximation to the value which the purity freezes, which is here
\begin{eqnarray} \label{heavy_freeze}
\gamma_{\bm{k}}^{\mfs}(\eta \to 0^{-}) \ \simeq \ 1 - 4 \pi^2 \epsilon f_1^2 e^{- 2 \pi \mu} \qquad \qquad \mathrm{when}\; \  \pi \mu = \pi \sqrt{ \frac{m^2}{H^2} - \frac{9}{4} } \gg 1\ . 
\end{eqnarray}
Notice that this quantity is independent of the scale $k$, and indeed tends to 1 as $m/H$ gets infinitely large. We show in Figure \ref{fig:freeze} that this formula is a good approximation for the late time value, so long as we are in the perturbative regime and $m/H \gtrsim 3/2$.

\section{Ekpyrotic decoherence}
\label{sec:contract}

In this section we consider decoherence in two-field ekpyrosis by studying the contracting phase with $\epsilon \geq \frac{3}{2}$ or $w\geq 0$ (encompassing cases (ii) and (iii) from \S\ref{sec:scaleinv}). In the cases where scale-invariant power spectra are produced, decoherence is observed to be a nearly ubiquitous phenomenon. The only exception arises when the coupling and isocurvature mass attain exceptionally small and large values (respectively). In such instances, the onset of decoherence is so late that it occurs near the Big Crunch and consequently falls beyond the scope of the domain of validity of the gravitational EFT.

\subsection{Purity loss for $\epsilon = \frac{3}{2}$}

Following case (ii) from \S\ref{sec:scaleinv}, we have here $\epsilon = \frac{3}{2}$ while also requiring 
\begin{eqnarray} \label{Delta_cond}
\Delta  \ = \  f_1^2 - \frac{1}{2} f_2 + \frac{1}{2} h_2 \  > \ 0\,,
\end{eqnarray}
so that the spectral index is $n_s=1$, where we have used the definition (\ref{spectral}) above. In what follows, figures are presented in terms of conformal time, with decoherence occurring as $\eta \to 0^{-}$ (the relation between the cosmic time and conformal time scales differently as in the inflationary case and there is not the same notion of $e$-folds in this case). 

\vspace{0.1in}

\begin{figure}[h]
\centering
\includegraphics[width=0.48\linewidth]{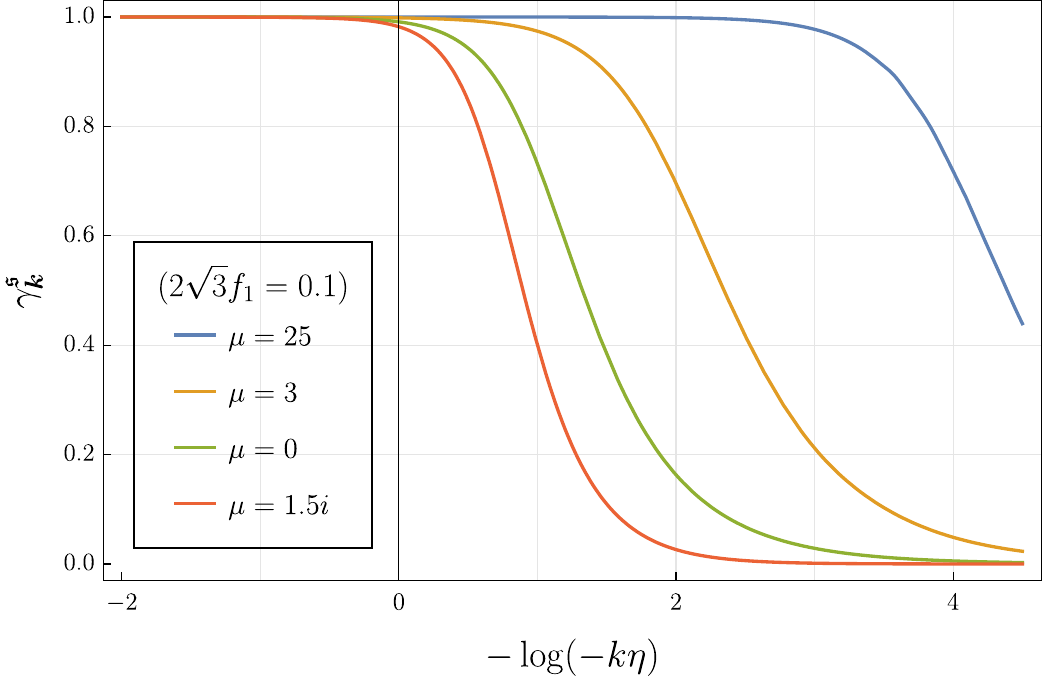} \ \ \includegraphics[width=0.48\linewidth]{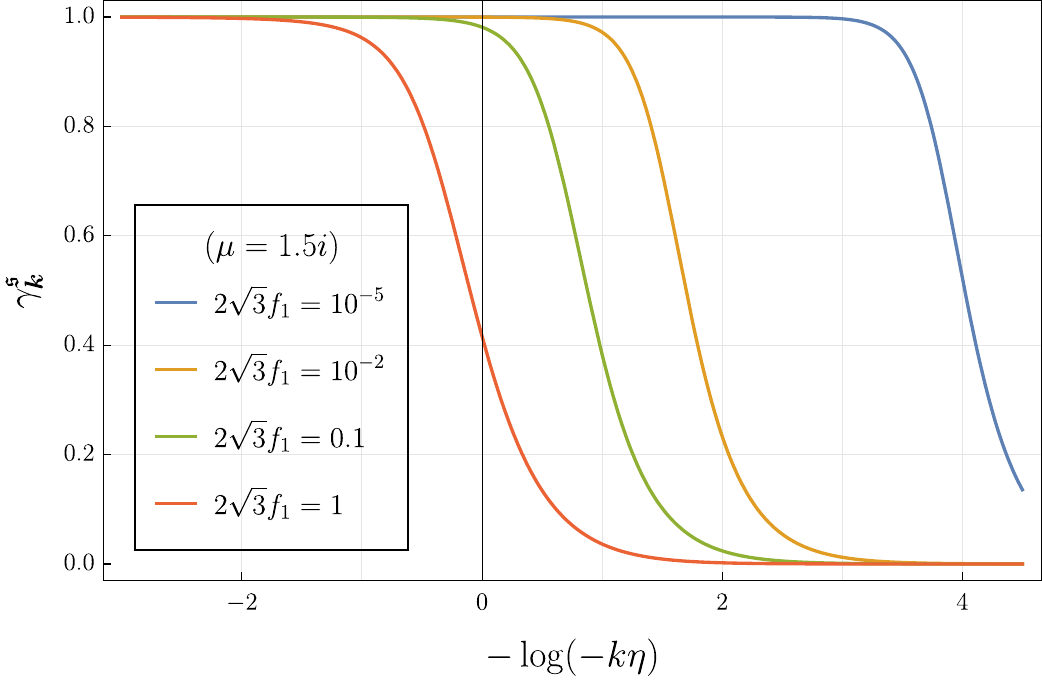}
\caption{Exact evolution of the purity $\gamma_{\bm{k}}^\mfs$ as a function of $-\log(-k \eta)$ (in the ekpyrotic case (ii) with $\epsilon = \frac{3}{2}$). Left panel: for various choices of $\mu$ (and fixed coupling $ 2 \sqrt{3} f_1 = 0.1$) Right panel: for various choices of coupling $f_1$ (and fixed $\mu = \frac{3i}{2}$). One finds faster decoherence for smaller masses and larger couplings. }
\label{figure:Fig8}
\end{figure}

As can be seen from Fig.~\ref{figure:Fig8}, the transition from a pure state to a mixed state is controlled by both $\mu$ and the coupling $f_1$. The larger the mass $\mu$ or the smaller the coupling $f_1$ the more time it takes for decoherence to take place --- this is notably different from the inflationary case (see {\it e.g.}~Figs.~\ref{figure:Fig1} and \ref{figure:Fig2}) in that - within the regime of validity of the EFT - we identify no recoherence for large masses in the ekpyrotic case, as we confirm below analytically. When the mass of the isocurvature mode is too large the onset of decoherence occurs beyond the regime of validity of the EFT and cannot be trusted. This is consistent with the expectation that for sufficiently large masses of the entropic mode, the latter effectively decouples and there is no mixing to enable decoherence to take place. Likewise, when the coupling parameter $f_1$ is small enough, the system remains predominantly pure throughout its evolution, with decoherence kicking in only at very late times. This means that there is always a choice of coupling $f_1$ which is so small that the purity remains at 1 for all times $\eta < \eta_{\mathrm{cut}}$ -- this can be understood as decoherence effectively kicking in {\it after} the time $\eta_{\mathrm{cut}}$ defined in Section~\ref{sec:cutoff}, at which point the field theory used to describe the system becomes unreliable. Note however that this requires dialling the coupling $f_1$ to be tremendously small, at which point higher-order interactions of the curvature fluctuations become important (and likely cause decoherence as well). Despite these considerations, the robustness of decoherence here remains evident.

\subsubsection{Comparison to perturbative treatment}

We again proceed in perturbation theory, with the $\epsilon = \frac{3}{2}$ system mode functions (\ref{pmodes_def}) simplifying to
\begin{eqnarray} \label{pmodes_5}
\mathrm{p}_{\bm{k}}(\eta) \ = \ - i \sqrt{ \frac{k}{2} } \; e^{ - i k \eta} \left( 1 - \frac{3i}{k\eta} - \frac{3}{(- k\eta)^2} \right) \ ,
\end{eqnarray}
meanwhile the form of the environment modes functions (\ref{xmodes_def}) remain unsimplified, with the parameter $\mu$ there taking the form
\begin{eqnarray}
\mu  \ =  \ \sqrt{ 12 ( h_2 - f_2 ) - \frac{9}{4} } \ . 
\end{eqnarray}
Notice that the condition $\Delta > 0$ in Eq.~(\ref{Delta_cond}) implies that $h_2 - f_2 > - 2 f_1^2$. Since the coupling should be small in the perturbative treatment this implies that the boundary of allowed values is $h_2 - f_2 \simeq 0$ and so $\mu = \frac{3}{2}i$, which means that $\mu$ can take on the usual values as in the inflationary case\footnote{{\it i.e.}~along the positive real line $\mu \in [0,\infty)$, as well as along the imaginary axis such that $\mu = i \nu$ with $0 \leq \nu < \frac{3}{2}$.}.

As before, one must compute the integral $I_{\bm{k}}(\eta)$ defined in (\ref{Ik_def}) in order to determine the purity in perturbation theory Eq.~(\ref{PP_ordered_2}) and its resummation Eq.~(\ref{purity_det_rel_2}). The computation is performed in Appendix \ref{App:Ik_ekpy}, where
\begin{eqnarray} \label{Iksq_body_ekpy}
\left| I_{\bm{k}}(\eta) \right|^2 \ = \ 6 f_1^2 \; \bigg| \; \frac{\pi \; \mathrm{sech}(\pi\mu) }{\sqrt{2}} - e^{- i \tfrac{\pi}{4}} \sqrt{ \frac{ - k \eta}{\pi}  } \; \big[ \; h_{\mu}(- k \eta) + h_{-\mu}(- k\eta) \; \big] \; \bigg|^2\,, \  \quad
\end{eqnarray}
with the definition
\begin{eqnarray}
h_{\mu}(z) & : = & \frac{ \Gamma(- i \mu) e^{ + \tfrac{\pi\mu}{2} }  }{1 + 2 i \mu}\; \left( \frac{z}{2} \right)^{i\mu} \ _2F_2 \;^{\frac{1}{2}+i\mu, \frac{1}{2}+i\mu}_{\frac{3}{2}+i\mu, 1 + 2 i \mu}(2 i z) \\
& \ & \;  +  \frac{6\pi e^{\tfrac{\pi\mu}{2}} \mathrm{csch}(\pi\mu) e^{i z}}{ 4 \mu^2 + 9 } \bigg[ - \left( 1 + \frac{i}{z} \right) J_{1 + i \mu}(z) + \bigg( - i + \frac{ \tfrac{3}{2} + i \mu }{z} \left( 1 + \frac{i}{z} \right)  \bigg) J_{i \mu}(z)  \bigg] \,,\notag 
\end{eqnarray}
where $_2F_{2}$ is a generalized hypergeometric function and $J$ a Bessel function of the first kind. One finds at early times, 
\begin{eqnarray}
| I_{\bm{k}}(\eta) |^2 \  \simeq \ f_1^2\bigg[ \frac{3}{4 (- k \eta)^2} + \mathcal{O}\Big( (- k \eta)^{-4} \Big) \bigg] \qquad \qquad \mathrm{for}\  - k \eta \gg 1 \ ,
\end{eqnarray}
which shows again that the purity slowly decreases from 1 at early times. Furthermore the late time $- k \eta \ll 1$ behaviour is given by (for any $\mu \neq \frac{3}{2}i$)
\begin{equation}\label{ekpyr_late}
| I_{\bm{k}}(\eta) |^2 \simeq \frac{54 f_1^2 }{(- k \eta)^3} \bigg[ \frac{ 2 \coth (\pi  \mu ) }{\mu  \left(4 \mu ^2+9\right)}  - \frac{ \Gamma (-i \mu )^2 }{\pi  (2 \mu +3 i)^2} \left(\frac{-k\eta}{2}\right)^{2 i \mu } - \frac{ \Gamma (i \mu )^2 }{\pi  (2 \mu - 3 i)^2} \left(\frac{-k\eta}{2}\right)^{-2 i \mu } \bigg] 
\end{equation}
which shows that the scaling is always {\it at least} $| I_{\bm{k}}(\eta) |^2 \propto (- k \eta)^{-3}$ or faster -- this allows us to conclude that decoherence always happens so long as $|I_{\bm{k}}(\eta)|^2 \gg 1$ for $\eta < \eta_{\mathrm{cut}}$, {\it i.e.}~so long as there is a period within the regime of validity of the effective field theory for which $|I_{\bm{k}}(\eta)|^2 \gg 1$. By inspection of Eq.~(\ref{ekpyr_late}), one can see that the condition $|I_{\bm{k}}(\eta)|^2 \gg 1$ fails when $f_1$ is too small or when $\mu$ is too big (as also demonstrated in the exact integration in Fig.~\ref{figure:Fig8}). First, for $\mu$ not too large, the decoherence condition $| I_{\bm{k}}(\eta) |^2  \gg 1$ implies that one needs
\begin{eqnarray} \label{deco_coupling_hier}
- k \eta_{\mathrm{cut}} \ll  -k \eta \ll f_1^{2/3} \ll 1 \ .
\end{eqnarray}
This constrains how small the coupling is allowed to be for decoherence to proceed -- in this case, one must have $f_{1}^2 \gg (- k \eta_{\mathrm{cut}})^3 \sim (- k \eta_0)^3 ( - \mpl \eta_0 )$. If $f_1$ is allowed to shrink to be of this size or smaller, higher order effect should be included and would in turn likely lead to a mixed state.

Alternatively, when $\mu \gg 1$ the condition (\ref{ekpyr_late}) takes the approximate form
\begin{equation}
| I_{\bm{k}}(\eta) |^2 \simeq \frac{27 f_1^2 }{(- k \eta)^3 \mu^3} \qquad \mathrm{when} - k \eta \ll 1\ \mathrm{and} \ \mu \gg 1 \ .
\end{equation} 
This expression is useful in that it highlights the hierarchy of scales required to decohere when $\mu \gg 1$, where one gets $|I_{\bm{k}}(\eta)|^2 \gg 1$ only when
\begin{eqnarray} \label{deco_mass_hier}
 - k \eta_{\mathrm{cut}} \ll  -k \eta \ll \frac{f_1^{2/3}}{\mu} \ll 1 \, .
\end{eqnarray}
This similarly shows that one requires $f_{1}^2 / \mu^3 \gg (- k \eta_{\mathrm{cut}})^3 \sim (- k \eta_0)^3 ( - \mpl \eta_0 )$ to have decoherence proceed.

\begin{figure}[h]
\centering
\includegraphics[width=0.96\linewidth]{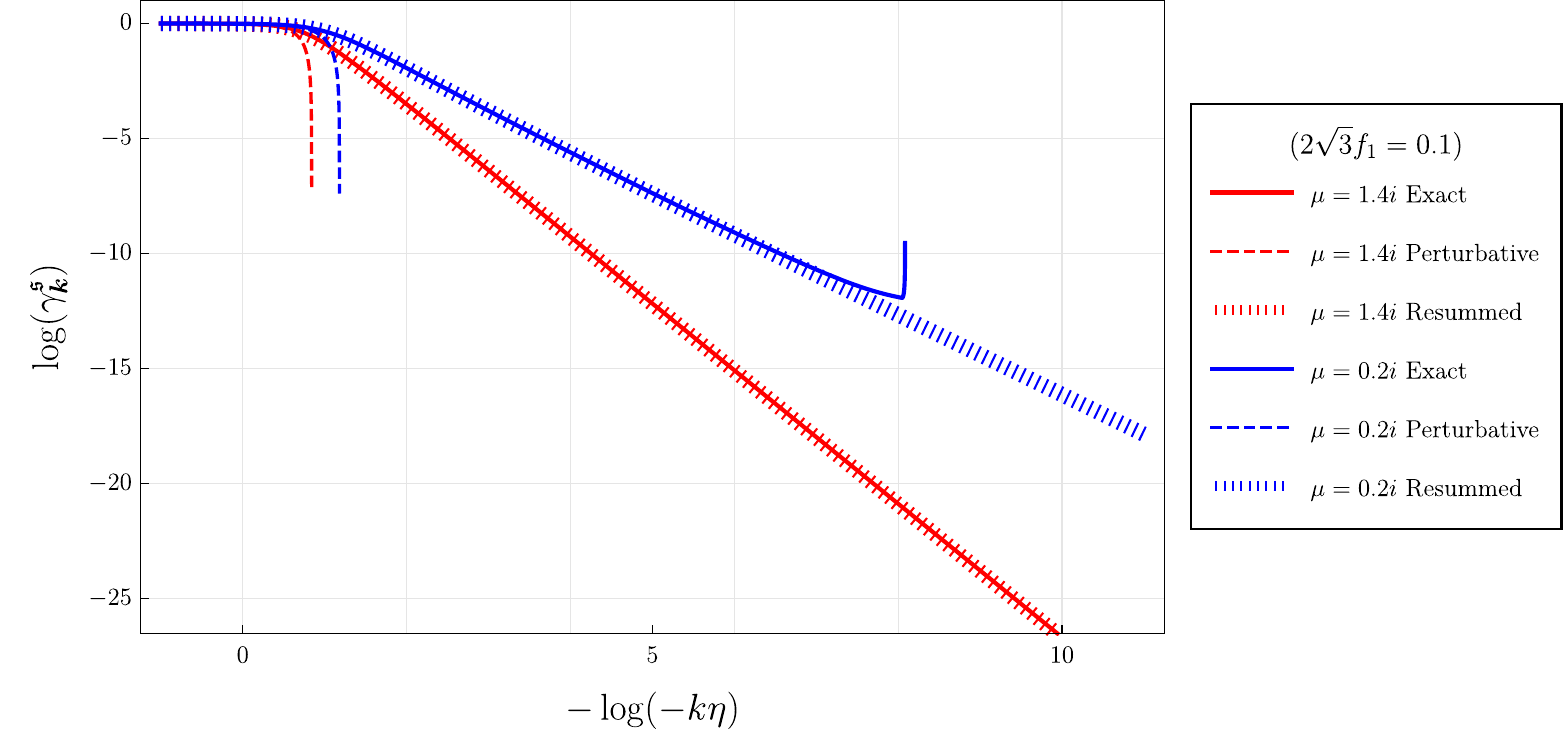}
\caption{Evolution of $\log( \gamma_{\bm{k}}^\mfs )$ as a function of $- \log( - k \eta)$ (in the ekpyrotic case (ii) with $\epsilon = \frac{3}{2}$) for coupling $2\sqrt{3} f_1 = 0.1$ as well as two choices of $\mu$. The exact evolution is compared against the perturbative and resummed evolution. Note that numerical noise starts to dominate the exact evolution for heavier masses at late times, at which point the exact numerical solution can no longer be trusted. }
\label{figure:Fig11}
\end{figure}

The scaling in the case $\mu = \frac{3}{2} i $ gives the fastest possible decoherence rate which is (see Eq.~(\ref{massless_ekpy_Ik}))
\begin{eqnarray} \label{massless_ekpy_body}
| I_{\bm{k}}(\eta) |^2 \simeq \frac{3 f_1^2 }{ (- k \eta)^6 } + \ldots \quad \mathrm{for}\  - k \eta \ll 1 \,.
\end{eqnarray}
In general, the scaling of $| I_{\bm{k}}(\eta) |^2$ decreases from $\sim (- k \eta)^{-6}$ to $\sim (- k \eta)^{-3}$ for $\mu = \frac{3i}{2}$ to $0$, and then for real values of $\mu$ the scaling remains at $(- k \eta)^{-3}$ with extra oscillatory behaviour (that is exponentially suppressed in $\mu$). Notice this behaviour in perturbation theory is in good agreement with the exact dynamics as shown in Figure \ref{figure:Fig11}. We conclude that (so long as $f_1^{2/3}, f_1^{2/3} / \mu \gg - k \eta_{\mathrm{cut}}$) the late time scaling of the purity is $\gamma_{\bm{k}}^{\mfs} \sim (- k \eta )^{\alpha} \ll 1$ with $\alpha$ varying between $3/2$ (heavy) and $3$ (massless).

\subsection{Purity loss for $\epsilon > \frac{3}{2}$}

For completeness we also consider case (iii) from \S\ref{sec:scaleinv}, which broadly agrees with the above behaviour albeit is a bit more complicated to deal with analytically. We have here $\epsilon > \frac{3}{2}$ with the additional condition $\Delta = \frac{3}{2} -\epsilon$ -- note that in this case Eq.~(\ref{full_tdepmass}) has the form
\begin{equation} 
M_{\mathrm{eff}}^2(\eta) + \lambda^2(\eta) \ = \ - \; \frac{4 \epsilon (\epsilon - \frac{3}{2} + f_1^2)}{( \epsilon - 1)^2\eta^2} \,,
\end{equation}
which implies that the total mass squared in the exact description is {\it negative} in this case.

\begin{figure}[h]
\centering
\includegraphics[width=0.49\linewidth]{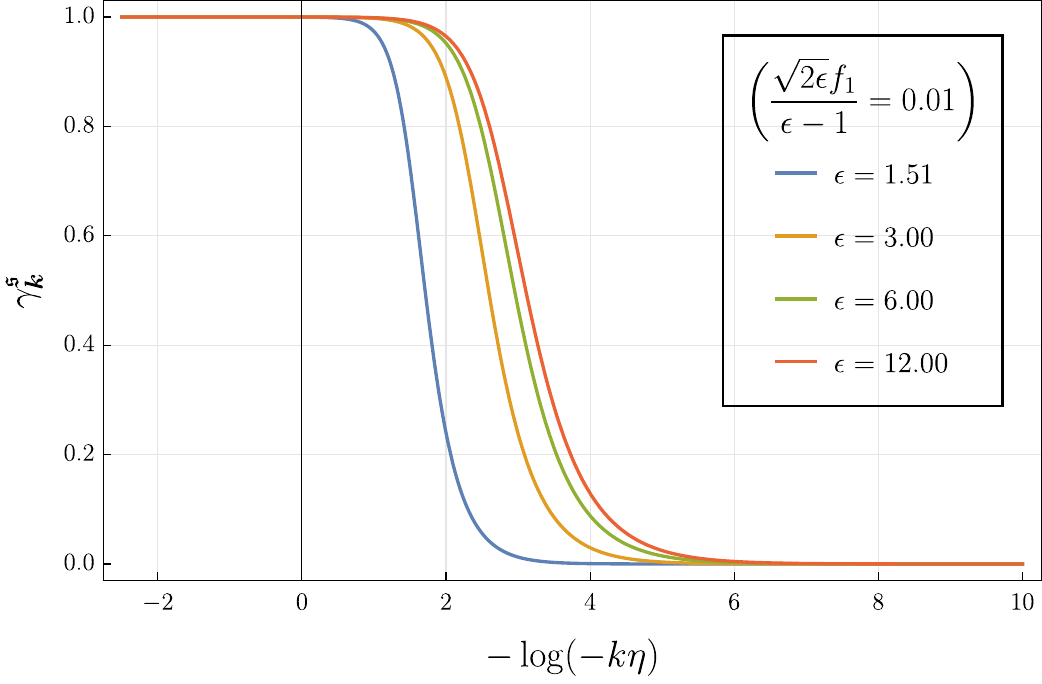}\ \ \includegraphics[width=0.49\linewidth]{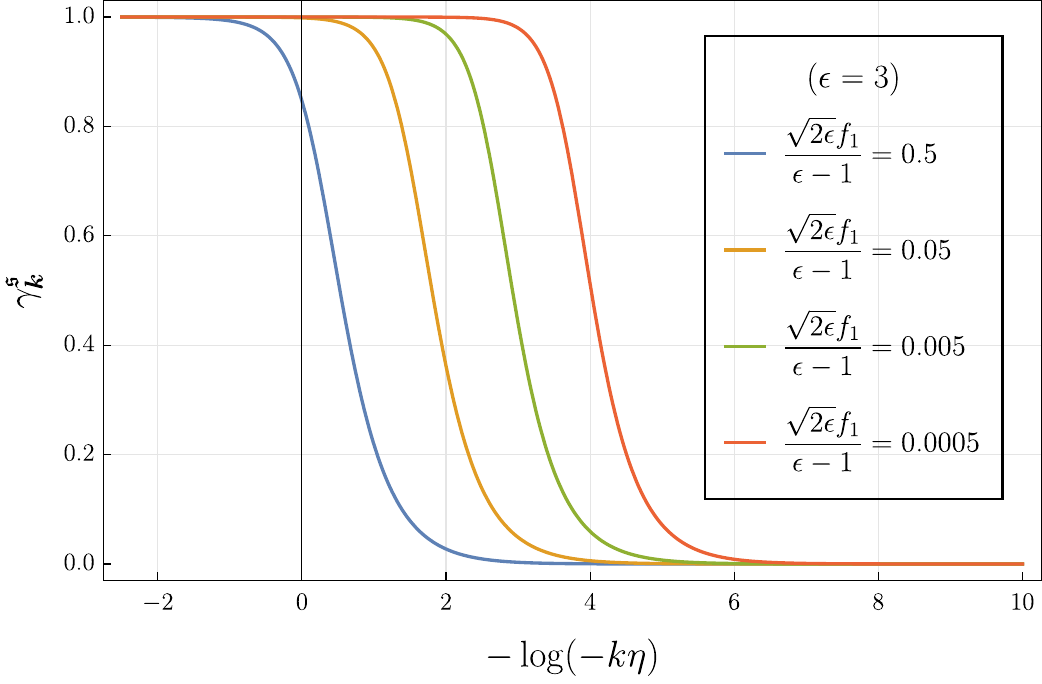}
\caption{Exact evolution of purity $\gamma_{\bm{k}}^\mfs$ as a function of $-\log(-k \eta)$ (in the ekpyrotic case (iii) with $\epsilon > \frac{3}{2}$). Left Panel: for various choices of $\epsilon$ (and for fixed $\frac{\sqrt{2\epsilon}}{\epsilon- 1} f_1 = 0.01$. Right Panel: for various choices of coupling (and for fixed $\epsilon = 3$. }
\label{figure:Fig12}
\end{figure}

The freedom in the parameters of the problem is more constrained here since the mass depends only on $\epsilon$ and $f_1$ -- in Figure~\ref{figure:Fig12} we study the exact description for various choices of $\epsilon$ and $f_1$. Once again decoherence in ubiquitous in this case and occurs the fastest for $\epsilon \to \frac{3}{2}$. Notice that smaller couplings delay the onset of decoherence which again implies that couplings cannot be arbitrarily small for decoherence to proceed.
The above behaviour can again be understood in perturbation theory, where one finds that $| I_{\bm{k}}(\eta)|^2$ in this case has the late-time form (see Eq.~(\ref{Ik_iii_App})),
\begin{eqnarray} \label{Iksq_iii}
|I_{\bm{k}}(\eta)|^2 & \simeq & f_1^2 \; \frac{\epsilon \Gamma^2(1 - \gamma) \Gamma^2(\nu) }{2 \pi^2 (\epsilon - 1)^2 (\gamma - \nu)^2}\; \left( \frac{- k \eta}{2} \right)^{2\gamma - 2\nu}  
\end{eqnarray}
where $\gamma$ is defined in (\ref{vmodes_def}) and the parameter $\nu$ is given by
\begin{eqnarray}
\nu  \ \simeq  \ \frac{\sqrt{ 9 - 30 \epsilon + 17 \epsilon^2 } }{2(\epsilon - 1)} + \mathcal{O}(f_1^2)\,,
\end{eqnarray}
to leading order in the coupling $f_1$ (see Eq.~(\ref{nu_def_app})). This means that the power with which the late-time limit $|I_{\bm{k}}(\eta)|^2 \propto (- k \eta)^{2\gamma - 2\nu}$ scales with is $\epsilon$-dependent such that
\begin{eqnarray}
2\gamma - 2\nu \ = \ 1 - \frac{ 2 + \sqrt{ 9 - 30 \epsilon + 17 \epsilon^2 } }{\epsilon - 1}\,,
\end{eqnarray}
where $2\gamma - 2\nu$ is a monotonically increasing function here, such that $2 \gamma - 2 \nu \to -6$ when $\epsilon \to \frac{3}{2}$, and $2 \gamma - 2 \nu \to - \sqrt{17} + 1 \simeq -3.12$ when $\epsilon \to \infty$. This confirms the qualitative behaviour in Figure~\ref{figure:Fig12}, where one finds that the fastest decoherence takes place for $\epsilon \to \frac{3}{2}^{+}$. Taking the square root, we again recover that the final value of the purity before the bounce is $\gamma_{\bm{k}}^{\mfs} \sim (- k \eta)^{\alpha} \ll 1$ where $\alpha$ varies between $\simeq 3/2$ ($\epsilon \rightarrow \infty$) and $3$ ($\epsilon \rightarrow \frac{3}{2}$).

\begin{figure}[h]
\centering
\includegraphics[width=0.96\linewidth]{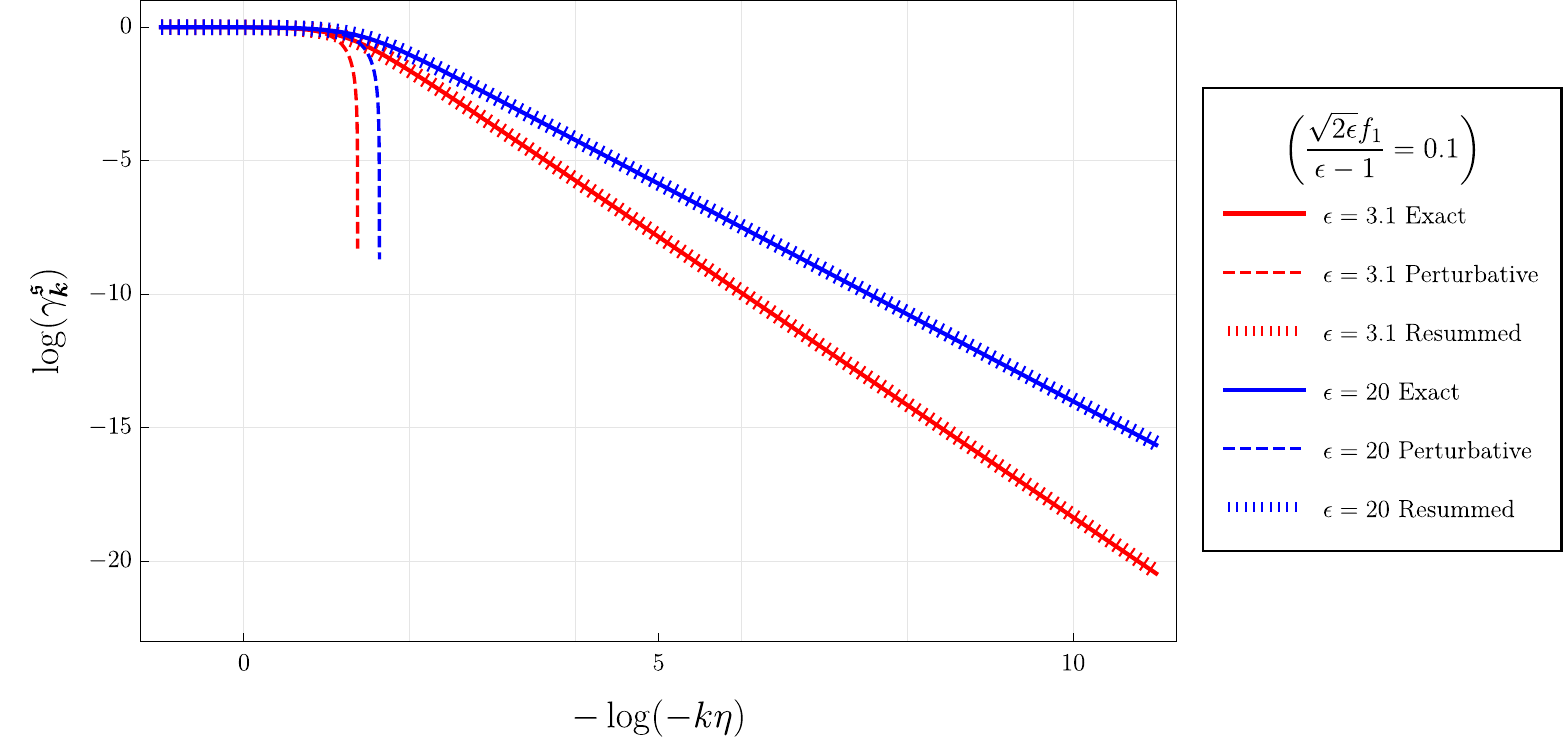}
\caption{Evolution of $\log( \gamma_{\bm{k}}^\mfs )$ as a function of $- \log( - k \eta)$ (in the ekpyrotic case (iii) with $\epsilon > \frac{3}{2}$) for coupling $\frac{\sqrt{2\epsilon}f_1}{\epsilon- 1} f_1 = 0.1$ as well as two choices of $\mu$. The exact evolution is compared against the perturbative and resummed evolution. }
\label{figure:Fig14}
\end{figure}

We confirm in Figure~\ref{figure:Fig14} that the perturbative treatment matches the exact evolution to a very good degree -- including up to very late times. 
Lastly, we comment on the vanishing/quenched off coupling case where decoherence does not occur -- to this end note that $|I_{\bm{k}}(\eta)|^2 \gg 1$ is satisfied roughly when $f_1^2 ( - k \eta)^{2 \gamma - 2 \nu} \gg 1$. Similar to case (ii), the value of $\eta_{\mathrm{cut}}$ defined in Section~\ref{sec:cutoff} places a constraint on the size of $f_1$ such that
\begin{eqnarray} \label{f1_ep_constraint}
f_1^2  \ \gg \ ( - k \eta_{\mathrm{cut}} )^{2\gamma - 2 \nu} \ \sim \ (- k \eta_0)^{1 - \tfrac{ 2 + \sqrt{ 9 - 30 \epsilon + 17 \epsilon^2 } }{\epsilon - 1}}\;  ( - \mpl \eta_0)^{-1 + \tfrac{3 + \sqrt{ 9 - 30 \epsilon + 17 \epsilon^2 } }{\epsilon} }  \ .
\end{eqnarray}
Morally this tells us again that $f_1$ cannot be arbitrarily small if decoherence is to take place -- otherwise for couplings satisfying (\ref{f1_ep_constraint}), the decoherence always occurs.

\section{Conclusions}
\label{sec:conc}

In this article, we emphasize the role played by the cosmological background in driving the quantum information properties of the quantum fluctuations seeding the late-time inhomogeneities observed in the universe. We consider in \S\ref{sec:Ekpy} a two-field model which accommodates for both expanding (inflation) and contracting (ekpyrosis) FLRW solutions generating a nearly scale invariant power spectrum for the curvature perturbations \cite{Tolley:2007nq} compatible with the current CMB \cite{Aghanim:2018eyx,Planck:2018jri} and LSS \cite{SDSS:2005xqv, BOSS:2014hwf,Colas:2019ret,DES:2022qpf} observations. In \S\ref{sec:EoM}, we implement an open-quantum-system treatment \cite{breuerTheoryOpenQuantum2002, Burgess:2022rdo, Colas:2023wxa} where the isocurvature modes, so far unobserved in cosmological experiments \cite{Planck:2018jri}, are integrated out. We examine the non-unitary reduced dynamics of curvature perturbations through an analysis of quantum decoherence, utilizing the purity as a measure of the mixedness of the state. In the perturbative regime, we use our simple perturbative resummation (\ref{purity_det_rel_2}) to circumvent any instances of secular growth, and are able to extract the late time scaling of the purity with remarkable accuracy as compared to the exact evolution (albeit a small window of mass regimes $m/H \gtrsim \mathcal{O}(1)$ in the inflationary case -- see Appendix \ref{App:MEcomparison}).
 
The inflationary results ($\epsilon \ll 1$) obtained in \cite{Colas:2022kfu} are recovered in \S\ref{sec:inflation}, where notably late time decoherence of the curvature perturbation crucially depends on the mass of the isocurvature modes (see Figure \ref{figure:Fig1}). In particular, for effective masses heavier than $3H/2$, recoherence takes place and the purity asymptotes a constant close to one at late-time with the deviation from perfect purity is controlled by a Boltzmann-factor suppressed quantity given in (\ref{heavy_freeze}). Decoupling is effective and the late-time dynamics may be approximated as unitary. This conclusion is in sharp contrast with the contracting case considered in \S\ref{sec:contract}. As depicted in Figure \ref{figure:Fig8}, with sufficient time, decoherence occurs regardless of the coupling or isocurvature mass during contracting phases. However, it is worth noting that there is a possibility that decoherence takes place so late that it approaches the time of the bounce, rendering the calculation unreliable -- in this case there is no decoherence within the domain of validity of the EFT. Decoherence proceeds at a much faster rate than the inflationary counterpart (see Table \ref{tab:summary}) and the quantum state of the system ends up in a highly mixed state when the bounce occurs and the EFT breaks down. 

As shown in \cite{Tolley:2007nq}, the model at hand is unstable for contracting phases and we also have no analytic control over the bounce, however this is secondary to the main point of our work which is to establish how different backgrounds can drastically change the outcome of decoherence. This study highlights the crucial role played by the background dynamics in the evolution of the quantum information properties of the quantum fluctuations amplified during the early universe. It establishes a formal analogy between primordial cosmology and driven-dissipative systems \cite{Martin:2007bw} from which techniques might be imported \cite{Burgess:2022rdo, Colas:2023wxa}.

A natural follow-up of this work would consist in the inclusion of higher-order non-linear interactions, though this task would require the extension of the currently used open-quantum-system techniques towards the study of non-Gaussian quantum states. It would be meaningful to perform a more careful comparison of our resummation to more standard master equation treatments (as briefly described in Appendix \ref{App:ME}). Also, relating the purity to other entropy measures such as the entanglement entropy would allow us to investigate if the rapid decoherence rate occurring in the contracting phase violates certain entropy bounds \cite{Deffner:2017cxz}.

Lastly, our work highlights that the near-decoupling limit in cosmology can have considerably richer behaviour than its flat space counterparts. It may therefore be interesting to more deeply explore the relationship between Open and Wilsonian EFTs in cosmological systems, and whether there exist useful non-unitary (and perhaps non-local \cite{Jazayeri:2022kjy,Jazayeri:2023xcj,Jazayeri:2023kji}) effective descriptions for more simply (and agnostically from a model-building point of view) studying such a limit. We leave this for future work.

\subsection*{Acknowledgements}

We would like to thank Cliff Burgess, Xingang Chen, Julien Grain, Rich Holman, Andrew J. Tolley and Vincent Vennin for useful discussions. This work has been supported by STFC consolidated grant ST/X001113/1, ST/T000791/1, ST/T000694/1 and ST/X000664/1. The work of CdR and GK at Imperial are also supported by a Simons Investigator award 690508.

\appendix

\section{$\mathrm{TCL}_{2}$ master equation}
\label{App:ME}

We devote this appendix to writing down the so-called $2^{\mathrm{nd}}$-order time convolutionless ($\mathrm{TCL}_{2}$) master equation for the dynamics of the reduced density matrix. The purpose of this is twofold: first to briefly show that the $\mathrm{TCL}_{2}$ master equation improves on perturbation theory (and the related resummation used in the main text) through study of the transport equations of the system and $\det \boldsymbol{\Sigma}_{\ssA}$. Secondly, the perturbative limit of the equation for $\det \boldsymbol{\Sigma}_{\ssA}$ gives the simple perturbative resummation used in Eq.~(\ref{purity_det_rel_2}) in the main text.

There are many sophisticated derivations of the so-called $\mathrm{TCL}_{2}$ master equation \cite{Lindblad:1975ef, breuerTheoryOpenQuantum2002, breuerTimelocalMasterEquations2002, whitneyStayingPositiveGoing2008, Scarlatella:2021vfr, Burgess:2022rdo, Colas:2022hlq}, but in essence it takes the result of perturbation theory like (\ref{PT_rhoA}) and extends its domain of validity by taking its time derivative and replacing
\begin{eqnarray} \label{ME_replacement}
\varrho_{\ssA}(\eta_{\mathrm{in}}) \simeq \varrho_{\ssA}(\eta)
\end{eqnarray}
on the RHS, giving the differential equation for the reduced density matrix:
\begin{eqnarray} \label{TCL2}
\frac{\mathcal{V}}{(2\pi)^3} \cdot \frac{\partial \varrho_{\ssA\bm{k}}^{\mfs}}{\partial \eta} & \simeq & \int_{\eta_{\mathrm{in}}}^{\eta} \exd \eta' \; \lambda(\eta) \lambda(\eta') \bigg( \mathcal{W}^{\mfs}_{\ssB\bm{k}}(\eta,\eta')  \big[ p^{\mfs}_{\bm{k}}(\eta') \varrho_{\ssA\bm{k}}^{\mfs}(\eta) ,  p^{\mfs}_{\bm{k}}(\eta) \big] \notag \\
& \ & \qquad  \qquad  \qquad  \qquad  \qquad \qquad + \mathcal{W}^{\mfs\ast}_{\ssB\bm{k}}(\eta,\eta')  \big[ p^{\mfs}_{\bm{k}}(\eta) , \varrho_{\ssA\bm{k}}^{\mfs}(\eta) p^{\mfs}_{\bm{k}}(\eta') \big] \bigg) + \mathcal{O}(\lambda^4) \,.
\end{eqnarray}
Note that the replacement (\ref{ME_replacement}) works when the initial state of the full density matrix is factorizable in the system and environment, as is the case here given the Bunch-Davies initial condition (\ref{BD_dec}). Furthermore this replacement must be justified at each time $\eta$ -- one way this can be done is by understanding (\ref{ME_replacement}) as a Taylor series like:
\begin{eqnarray} \label{ME_replacement2}
\varrho_{\ssA}(\eta_{\mathrm{in}}) \simeq \varrho_{\ssA}(\eta) + (\eta_{\mathrm{in}} - \eta) \partial_{\eta} \varrho_{\ssA}(\eta) + \ldots
\end{eqnarray}
The sub-leading terms are all proportional to time derivatives of the reduced density matrix and these are dropped typically on the grounds of weak couplings ({\it i.e.}~see (\ref{TCL2}) which shows $\varrho'_{\ssA} \sim \mathcal{O}(\lambda^2$)). Since these terms are also proportional to powers of $\eta_{\mathrm{in}} - \eta$, these can clearly be in danger of growing parametrically large compared to the leading term of (\ref{ME_replacement2}). This suggests that master equations improve upon straight-up perturbation theory at late times, but can also fail at even later times.

One may write (\ref{TCL2}) in a clearer form, by expressing the operator $p^{\mfs}_{\bm{k}}(\eta')$ in terms of system operators at the time $\eta$ where
\begin{eqnarray} \label{etap_to_eta}
p^{\mfs}_{\bm{k}}(\eta') \ = \ - 2 \mathrm{Im}[ \mathrm{p}_{\bm{k}}(\eta) \mathrm{p}^{\ast}_{\bm{k}}(\eta')  ] \; v^{\mfs}_{\bm{k}}(\eta) + 2 \mathrm{Im}[ \mathrm{v}_{\bm{k}}(\eta) \,,\mathrm{p}^{\ast}_{\bm{k}}(\eta') ] \; p^{\mfs}_{\bm{k}}(\eta)
\end{eqnarray}
which is derived using the Wronskian relation $\mathrm{v}_{\bm{k}}(\eta) \mathrm{p}^{\ast}_{\bm{k}}(\eta')- \mathrm{v}^{\ast}_{\bm{k}}(\eta) \mathrm{p}^{\ast}_{\bm{k}}(\eta) = + i $. This turns (\ref{TCL2}) into
\begin{eqnarray} \label{TCL2_2}
\frac{\mathcal{V}}{(2\pi)^3} \cdot \frac{\partial \varrho_{\ssA\bm{k}}^{\mfs}}{\partial \eta} & \simeq & - \mathcal{A}_{\bm{k}}(\eta) [ v^{\mfs}_{\bm{k}}(\eta) \varrho^{\mfs}_{\ssA\bm{k}}(\eta), p^{\mfs}_{\bm{k}}(\eta) ] -  \mathcal{A}^{\ast}_{\bm{k}}(\eta) [ p^{\mfs}_{\bm{k}}(\eta), \varrho_{\bm{k}}(\eta) v^{\mfs}_{\bm{k}}(\eta)  ]  \\
& \ & \qquad +  \mathcal{B}_{\bm{k}}(\eta) [ p^{\mfs}_{\bm{k}}(\eta) \varrho^\mfs_{\ssA\bm{k}}(\eta), p^{\mfs}_{\bm{k}}(\eta) ] + \mathcal{B}^{\ast}_{\bm{k}}(\eta) [ p^{\mfs}_{\bm{k}}(\eta), \varrho^\mfs_{\ssA\bm{k}}(\eta) p^{\mfs}_{\bm{k}}(\eta)  ] \,, \notag
\end{eqnarray}
with the definitions
\begin{eqnarray}
\begin{aligned} \label{AB_defs}
\mathcal{A}_{\bm{k}}(\eta) & := \int_{\eta_{\mathrm{in}}}^\eta \exd \eta' \; 2 \lambda(\eta) \lambda(\eta') \mathrm{Im}\big[ \mathrm{p}_{\bm{k}}(\eta) \mathrm{p}^{\ast}_{\bm{k}}(\eta') \big] \mathrm{x}_{\bm{k}}(\eta) \mathrm{x}^{\ast}_{\bm{k}}(\eta')  \\
\mathcal{B}_{\bm{k}}(\eta) & :=  \int_{\eta_{\mathrm{in}}}^\eta \exd s \; 2 \lambda(\eta) \lambda(\eta') \mathrm{Im} \big[ \mathrm{v}_{\bm{k}}(\eta) \mathrm{p}^{\ast}_{\bm{k}}(s) \big] \mathrm{x}_{\bm{k}}(\eta) \mathrm{x}^{\ast}_{\bm{k}}(s)
\end{aligned} 
\end{eqnarray}
where we have made use of the above relation (\ref{W_modes}). 

Armed with this master equation we derive a set of effective transport equations\footnote{To derive the effective transport equations resulting from the $\mathrm{TCL}_2$ master equation (\ref{TCL2_2}), one differentiates the definition (\ref{corr_Eq_def}) of $\Sigma_{\bm{k},ij}^{\mfs}$ and uses the equations of motion (\ref{Heis_EoM}). One also uses the master equation to write 
\begin{eqnarray}
\frac{\mathcal{V}}{(2\pi)^3} \cdot \mathrm{Tr}\left[ O^{\mfs}_{\bm{k}} \partial_\eta \varrho^{\mfs}_{\ssA\bm{k}} \right] & = & - \mathcal{A}_{\bm{k}} \mathrm{Tr}\left( \left[ p^{\mfs}_{\bm{k}} , O^\mfs_{\bm{k}} \right] v^{\mfs}_{\bm{k}}  \varrho^{\mfs}_{\ssA\bm{k}} \right) + \mathcal{B}_{\bm{k}} \mathrm{Tr}\left( \left[ p^\mfs_{\bm{k}} , O^\mfs_{\bm{k}} \right] p^\mfs_{\bm{k}} \varrho^\mfs_{\ssA\bm{k}} \right) + \mathrm{H.c.}
\end{eqnarray}
and uses the commutators $[ p^{\mfs}_{\bm{k}} , v^\mfs_{\bm{k}} v^\mfs_{\bm{k}} ] = - 2 i \Scale[0.85]{ \delta^{(3)}(\bm{0})} v^\mfs_{\bm{k}}$, $[ p^{\mfs}_{\bm{k}} , \tfrac{\{ v^\mfs_{\bm{k}} , p^{\mfs}_{\bm{k}} \}}{2}  ]=  - i \Scale[0.85]{ \delta^{(3)}(\bm{0})} p^{\mfs}_{\bm{k}}$ and $[ p^{\mfs}_{\bm{k}} , p^{\mfs}_{\bm{k}} p^{\mfs}_{\bm{k}} ] = 0$.}  for the correlators of the open system $\Sigma_{\bm{k}, 11}^{\mfs}$, $\Sigma_{\bm{k},12}^{\mfs}$ and $\Sigma_{\bm{k},22}^{\mfs}$ from Eq.~(\ref{system_corr_mat}) where 
\begin{equation}
\frac{\partial}{\partial \eta} \left[ \begin{matrix} \Sigma^\mfs_{\bm{k},11} \\ \Sigma^\mfs_{\bm{k},12} \\ \Sigma^\mfs_{\bm{k},22} \end{matrix} \right] \bigg|_{\mathrm{TCL}_2} \ \simeq \ \left[
\begin{array}{ccc}
 \frac{2}{(\epsilon -1)\eta} - 4 \mathrm{Im}\left[ \mathcal{A}_{\bm{k}} \right] & 2 + 4 \mathrm{Im}\left[ \mathcal{B}_{\bm{k}} \right] & 0 \\
 - k^2 & - 2 \mathrm{Im}\left[ \mathcal{A}_{\bm{k}} \right]  & 1 + 2 \mathrm{Im}\left[ \mathcal{B}_{\bm{k}} \right]  \\
 0 & -2 k^2 & -\frac{2}{(\epsilon -1)\eta } \\
\end{array}
\right] \left[ \begin{matrix} \Sigma^\mfs_{\bm{k},11} \\ \Sigma^\mfs_{\bm{k},12} \\ \Sigma^\mfs_{\bm{k},22} \end{matrix} \right]  + \left[ \begin{matrix} 2 \mathrm{Re}\left[ \mathcal{B}_{\bm{k}} \right] \\ \mathrm{Re}\left[ \mathcal{A}_{\bm{k}} \right]  \\ 0  \end{matrix} \right]  \ ,
\end{equation}
{\it c.f.}~the exact transport equations in Eq.~(\ref{Transport_exact}). This also predicts the effective $\mathrm{TCL}_{2}$ equation
\begin{eqnarray} \label{TCL2_det}
\frac{\partial \det \boldsymbol{\Sigma}^{\mfs}_{\ssA\bm{k}} }{\partial \eta} \bigg|_{\mathrm{TCL}_2} \ \simeq \ - 4 \mathrm{Im}[\mathcal{A}_{\bm{k}}] \; \det \boldsymbol{\Sigma}^{\mfs}_{\ssA\bm{k}} - 2 \mathrm{Re}[\mathcal{A}_{\bm{k}}] \; \Sigma^{\mfs}_{\bm{k},12} + 2 \mathrm{Re}[\mathcal{B}_{\bm{k}}] \; \Sigma^{\mfs}_{22,\bm{k}}   \ .
\end{eqnarray}
{\it c.f.}~the exact equation of motion for $\det \boldsymbol{\Sigma}^{\mfs}_{\ssA\bm{k}}$ in Eq.~(\ref{det_exact}).

\subsection{Perturbative limit}
\label{App:det_PT}

Here we consider the perturbative limit of the effective equation for $\det \boldsymbol{\Sigma}^{\mfs}_{\ssA\bm{k}}$. This amounts to inserting the initial state $\rho_{\ssA}(\eta_{\mathrm{in}})$ (as opposed to $\rho_{\ssA}(\eta)$) on the RHS of Eq.~(\ref{TCL2_det}) above -- this turns out to mean $\Sigma^\mfs_{\bm{k},12} \simeq \mathrm{Re}\left[ \mathrm{v}_{\bm{k}}(\eta) \mathrm{p}^{\ast}_{\bm{k}}(\eta) \right]$ as well $\Sigma^\mfs_{\bm{k},22} \simeq | \mathrm{p}_{\bm{k}}(\eta) |^2$ as well as $\boldsymbol{\Sigma}^{\mfs}_{\ssA\bm{k}} \simeq \frac{1}{4}$ on the RHS. With this (\ref{TCL2_det}) becomes
\begin{eqnarray}
\frac{\partial \det \boldsymbol{\Sigma}^{\mfs}_{\ssA\bm{k}} }{\partial \eta} \; \bigg|_{\mathrm{PT}} & \simeq & - \mathrm{Im}[\mathcal{A}_{\bm{k}}]  - 2 \mathrm{Re}\left[ \mathrm{v}_{\bm{k}}(\eta) \mathrm{p}^{\ast}_{\bm{k}}(\eta) \right] \; \mathrm{Re}[\mathcal{A}_{\bm{k}}] + 2 | \mathrm{p}_{\bm{k}}(\eta) |^2  \mathrm{Re}[\mathcal{B}_{\bm{k}}]  \\
& = & \int_{\eta_{\mathrm{in}}}^\eta \exd \eta' \; 2 \lambda(\eta) \lambda(\eta') \bigg( \Scale[0.85]{  - \mathrm{Im}\big[ \mathrm{p}_{\bm{k}}(\eta) \mathrm{p}^{\ast}_{\bm{k}}(\eta') \big] \mathrm{Im}[ \mathrm{x}_{\bm{k}}(\eta) \mathrm{x}^{\ast}_{\bm{k}}(\eta') ] } \\
&\ & \qquad \quad \Scale[0.85]{ + \bigg\{ - 2 \mathrm{Re}\left[ \mathrm{v}_{\bm{k}}(\eta) \mathrm{p}^{\ast}_{\bm{k}}(\eta) \right]  \mathrm{Im}\big[ \mathrm{p}_{\bm{k}}(\eta) \mathrm{p}^{\ast}_{\bm{k}}(\eta') \big] + 2 | \mathrm{p}_{\bm{k}}(\eta) |^2 \mathrm{Im} \big[ \mathrm{v}_{\bm{k}}(\eta) \mathrm{p}^{\ast}_{\bm{k}}(s) \big] \bigg\}  \mathrm{Re}[\mathrm{x}_{\bm{k}}(\eta) \mathrm{x}^{\ast}_{\bm{k}}(s) ] } \bigg) \notag
\end{eqnarray}
where we've also used (\ref{AB_defs}). Using the Wronskian (see the text below Eq.~(\ref{etap_to_eta})) one finds that the quantity in the curly brackets simplifies to $\mathrm{Re}[\mathrm{p}_{\bm{k}}(\eta) \mathrm{p}^{\ast}_{\bm{k}}(\eta')]$ reducing the above to
\begin{eqnarray}
\frac{\partial \det \boldsymbol{\Sigma}^{\mfs}_{\ssA\bm{k}} }{\partial \eta} \; \bigg|_{\mathrm{PT}} & \simeq & \int_{\eta_{\mathrm{in}}}^\eta \exd \eta' \; \Scale[0.85]{ 2 \lambda(\eta) \lambda(\eta') \bigg(   - \mathrm{Im}\big[ \mathrm{p}_{\bm{k}}(\eta) \mathrm{p}^{\ast}_{\bm{k}}(\eta') \big] \mathrm{Im}[ \mathrm{x}_{\bm{k}}(\eta) \mathrm{x}^{\ast}_{\bm{k}}(\eta') ] + \mathrm{Re}\big[ \mathrm{p}_{\bm{k}}(\eta) \mathrm{p}^{\ast}_{\bm{k}}(\eta') \big] \mathrm{Re}[\mathrm{x}_{\bm{k}}(\eta) \mathrm{x}^{\ast}_{\bm{k}}(s) ] \bigg) } \notag \\
& = & 2 \int_{\eta_{\mathrm{in}}}^\eta \exd \eta' \; \lambda(\eta) \lambda(\eta') \mathrm{Re} \big[ \mathrm{p}_{\bm{k}}(\eta) \mathrm{p}^{\ast}_{\bm{k}}(\eta') \mathrm{x}_{\bm{k}}(\eta) \mathrm{x}^{\ast}_{\bm{k}}(\eta') \big]\,.
\end{eqnarray}
Integrating the above with respect to $\eta$ (and using the initial condition $\det \boldsymbol{\Sigma}^{\mfs}_{\ssA\bm{k}}(\eta_{\mathrm{in}}) = \frac{1}{4}$) one finds that 
\begin{equation}
\det \boldsymbol{\Sigma}^{\mfs}_{\ssA\bm{k}}(\eta)  \big|_{\mathrm{PT}}  \simeq  \frac{1}{4} +  2 \int_{\eta_{\mathrm{in}}}^\eta \exd \eta_1 \int_{\eta_{\mathrm{in}}}^{\eta_1} \exd \eta_2 \; \lambda(\eta_1) \lambda(\eta_2) \mathrm{Re} \big[ \mathrm{p}_{\bm{k}}(\eta_1) \mathrm{p}^{\ast}_{\bm{k}}(\eta_2) \mathrm{x}_{\bm{k}}(\eta_1) \mathrm{x}^{\ast}_{\bm{k}}(\eta_2) \big] \,.
\end{equation}
This time-ordered double-integral can be simplified by expanding out the real part and then relabelling $\eta_1 \leftrightarrow \eta_2$ in the second term so that
\begin{eqnarray}
\det \boldsymbol{\Sigma}^{\mfs}_{\ssA\bm{k}}(\eta)  \big|_{\mathrm{PT}} & \simeq & \frac{1}{4} +  \int_{\eta_{\mathrm{in}}}^\eta \exd \eta_1 \int_{\eta_{\mathrm{in}}}^{\eta_1} \exd \eta_2 \; \lambda(\eta_1) \lambda(\eta_2) \mathrm{Re} \big[ \mathrm{p}_{\bm{k}}(\eta_1) \mathrm{p}^{\ast}_{\bm{k}}(\eta_2) \mathrm{x}_{\bm{k}}(\eta_1) \mathrm{x}^{\ast}_{\bm{k}}(\eta_2) \big] \\
&& + \int_{\eta_{\mathrm{in}}}^\eta \exd \eta_2 \int_{\eta_{\mathrm{in}}}^{\eta_2} \exd \eta_1 \; \lambda(\eta_2) \lambda(\eta_1)  \mathrm{p}^{\ast}_{\bm{k}}(\eta_2) \mathrm{p}_{\bm{k}}(\eta_1) \mathrm{x}^{\ast}_{\bm{k}}(\eta_2) \mathrm{x}_{\bm{k}}(\eta_1) \, . \notag
\end{eqnarray}
The integrands of both terms are then the same, with the first term integrating over the isosceles triangle below the line $\eta_2 = \eta_1$ in the $(\eta_1,\eta_2)$-plane, and the latter term integrating over the isosceles triangle above the same line --- the sum of the two terms then results in an unnested double-integral over the square $[0,\eta]^2$ (as in Eq.~(\ref{PP_ordered}) in the main text) such that
\begin{eqnarray}
\det \boldsymbol{\Sigma}^{\mfs}_{\ssA\bm{k}}(\eta)  \big|_{\mathrm{PT}} & \simeq & \frac{1}{4} +  \int_{\eta_{\mathrm{in}}}^\eta \exd \eta_1 \int_{\eta_{\mathrm{in}}}^\eta \exd \eta_2 \; \lambda(\eta_1) \lambda(\eta_2) \mathrm{p}_{\bm{k}}(\eta_1) \mathrm{p}^{\ast}_{\bm{k}}(\eta_2) \mathrm{x}_{\bm{k}}(\eta_1) \mathrm{x}^{\ast}_{\bm{k}}(\eta_2) \\
& = & \frac{1}{4} + \big| I_{\bm{k}}(\eta) \big|^2 \label{det_Pert_App}
\end{eqnarray}
with $I_{\bm{k}}(\eta)$ defined in (\ref{Ik_def}). This gives the perturbative purity resummation. We note in passing that one may also arrive at this result by using the perturbative limit of the RHS of the exact evolution of $\det \boldsymbol{\Sigma}^{\mfs}_{\ssA\bm{k}}$ in Eq.~(\ref{det_exact}).

\subsection{Resummation schemes}
\label{App:MEcomparison}

For completeness, we here briefly demonstrate the effectiveness of the master equation at resumming late time evolution, as compared to perturbation theory and the simple resummation used in the main text. This is most apparent in the regime where the perturbative resummation Eq.~(\ref{purity_det_rel_2}) fails to correctly capture the late time behaviour of the purity: this is the intermediate mass regime where $m/H \gtrsim 1$ (but not so large that one approaches the decoupling limit).

\begin{figure}[h]
\centering
  \includegraphics[width=0.96\linewidth]{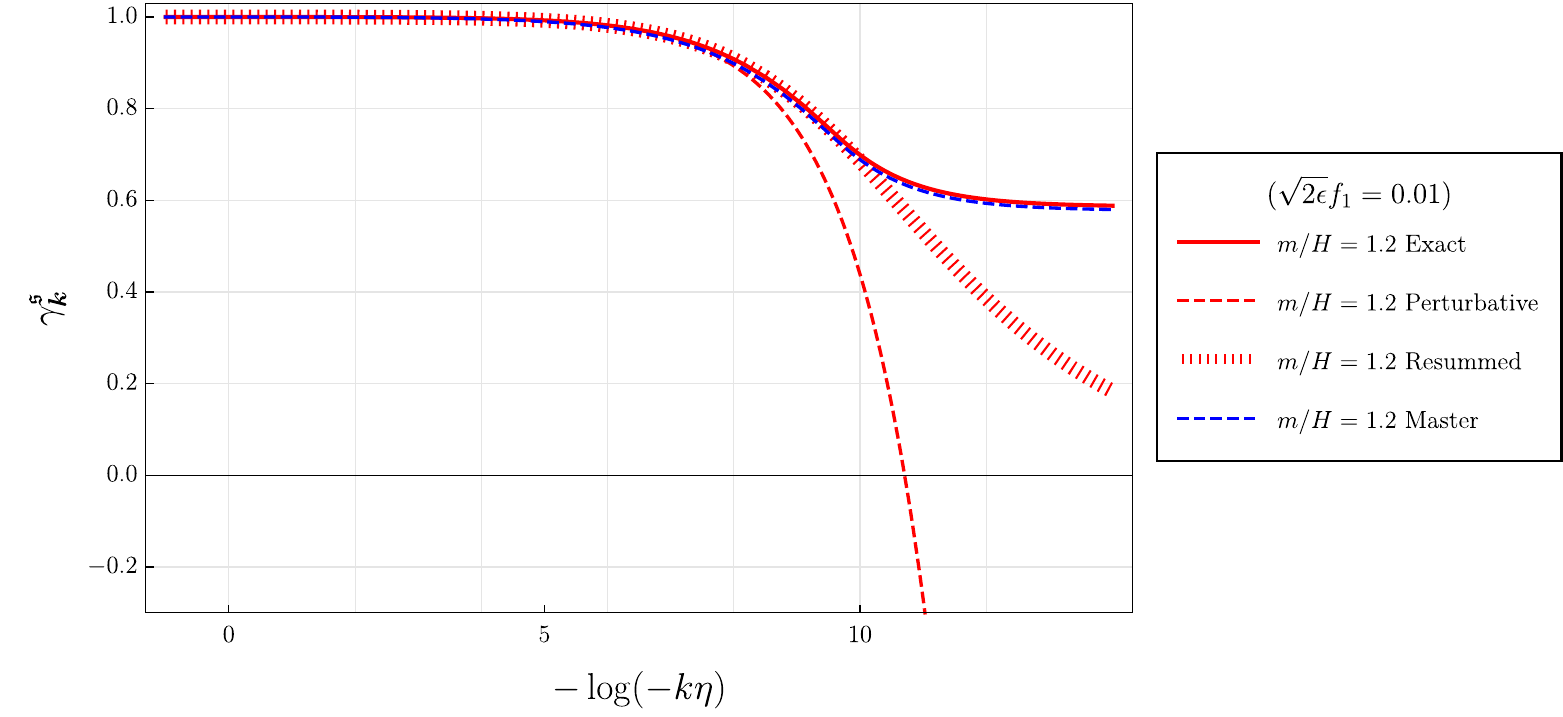} 
\caption{Exact vs.~perturbative vs. resummed vs.~$\mathrm{TCL}_{2}$ master equation evolution for the purity $\gamma_{\bm{k}}^\mfs$ as a function of cosmic time (in the inflationary case) for the choice $m/H = 1.2$ and $\sqrt{2 \epsilon} f_1 = 0.01$. }
\label{figure:App_ME}
\end{figure}

In Figure~\ref{figure:App_ME} we integrate the master equation along the lines of \cite{Colas:2022kfu}, in particular by neglecting ``spurious terms'' depending on the initial time $\eta_{\mathrm{in}}$ so that the appropriate late-time resummation is captured\footnote{This means dropping the $\eta_{\mathrm{in}}$-dependence of the $\mathrm{TCL}_{2}$ coefficients $\mathcal{A}_{\bm{k}}$ and $\mathcal{B}_{\bm{k}}$. See \cite{Colas:2022hlq,Colas:2022kfu} for more details on the precise procedure.}.

In this regime with $m/H \gtrsim 1$, the final value of the purity has a highly non-trivial dependence on the parameters in the problem. As one can see from Figure \ref{figure:App_ME}, the master equation evolution follows the exact evolution well (with some small error) -- as found in \cite{Colas:2022kfu}, the late-time value of the purity is difficult to attain analytically (although one may generally find the rate at which the purity approaches the final value from the master equation). Meanwhile the perturbative resummation does a poor job of predicting the late-time value, since its growth $| I_{\bm{k}}(\eta)|^2 \propto (-k\eta)^{1+ 2i \mu}$ is so slow, that $| I_{\bm{k}}(\eta)|^2 \gg 1$ never becomes appreciably true before higher-order effects in the coupling become important.

We close off this appendix by noting that the general problem of assessing the relative sizes of the error of the $\mathrm{TCL}_2$ master equation vs.~the perturbative resummation in the other mass regimes ({\it i.e.}~where the perturbative resummation does well at approximating the purity) is a more challenging problem to quantify, and we leave this for study in future work.

\section{Perturbative purity integrals}

In this Appendix we provide the details of the integration of $I_{\bm{k}}(\eta)$ defined in (\ref{Ik_def}), repeated here
\begin{eqnarray} \label{Ik_def_App}
I_{\bm{k}}(\eta) & := & \int_{-\infty- i \epsilon}^{\eta} \exd \eta' \; \lambda(\eta') \mathrm{p}_{\bm{k}}(\eta') \mathrm{x}_{\bm{k}}(\eta')\,,
\end{eqnarray}
which appears in the perturbative purity as $\gamma_{\bm{k}}^{\mfs}(\eta) \simeq 1 - 2 \big| I_{\bm{k}}(\eta) \big|^2$ in Eq.~(\ref{PP_ordered_2}). We assume $\eta_{\mathrm{in}} \to -\infty - i \epsilon$ here.

\subsection{Inflationary case}
\label{App:Ik_inflation}

Using the mode functions (\ref{pmodes_def}) and (\ref{xmodes_def}) as well as the coupling (\ref{coupling_def}) in the inflationary case with $\epsilon \ll 1$, the formula (\ref{Ik_def_App}) becomes
\begin{eqnarray}
I_{\bm{k}}(\eta) = - i f_1 \sqrt{ \frac{\pi \epsilon}{4} }\; e^{-  \tfrac{\pi}{2}\mu + i \tfrac{\pi}{4}} \int_{-k \eta}^{\infty} \exd z' \;  \frac{ e^{i z'} H^{(1)}_{i \mu}(z') }{ \sqrt{ z' }  } \,,
\end{eqnarray}
where we have used the integration variable $z' = - k \eta'$. It turns out that one can write down the exact primitive for the above $z'$ integral:
\begin{eqnarray}
\int \exd z' \; \frac{ e^{ i z'} H_{i \mu}^{(1)}(z') }{\sqrt{z'}} = - \frac{ 2 i e^{ + \tfrac{\pi\mu}{2} }  \sqrt{z'} }{ \pi } \; \bigg( \; g_{\mu}(z') \; + \; g_{-\mu}(z') \; \bigg)
\end{eqnarray}
where we define for convenience
\begin{equation} \label{gmu_def_App}
g_{\mu}(z') := \frac{ \Gamma(- i \mu) e^{ + \tfrac{\pi\mu}{2} }  }{1 + 2 i \mu}\; \left( \frac{z'}{2} \right)^{i\mu} \ _2F_2 \;^{\frac{1}{2}+i\mu, \frac{1}{2}+i\mu}_{\frac{3}{2}+i\mu, 1 + 2 i \mu}(2 i z')\,,
\end{equation}
where $_2F_{2}$ is a generalized hypergeometric function. Evaluating the above primitive at the end points results in
\begin{eqnarray}
I_{\bm{k}}(\eta) = - i \sqrt{\epsilon}  f_1 \; \bigg( \; \frac{\pi \; \mathrm{sech}(\pi\mu) }{\sqrt{2}} - e^{- i \tfrac{\pi}{4}} \sqrt{ \frac{ - k \eta}{\pi}  } \; \big[ \; g_{\mu}(- k \eta) + g_{-\mu}(- k\eta) \; \big] \; \bigg) \ . \quad
\end{eqnarray}
The late-time limit $- k \eta \ll 1$ of this function takes the form
\begin{eqnarray}
I_{\bm{k}}(\eta) & \simeq & - i \sqrt{\frac{\epsilon}{2}} \; f_1 \; \bigg( \; \pi \mathrm{sech}(\pi\mu) -\frac{ \Gamma (-i \mu ) e^{-\tfrac{i \pi }{2} \left(\tfrac{1}{2}+i \mu \right)}}{\sqrt{\pi } \left(\frac{1}{2}+i \mu \right)} \left(\frac{-k\eta}{2}\right)^{\frac{1}{2}+i \mu } \big[ 1 + \mathcal{O}(- k \eta) \big] \\
& \ & - \frac{ \Gamma (i \mu ) e^{-\tfrac{i \pi }{2} \left(\tfrac{1}{2}-i \mu \right)} }{\sqrt{\pi } \left(\frac{1}{2}-i \mu \right)} \left(\frac{-k\eta}{2}\right)^{\frac{1}{2}-i \mu } \big[ 1 + \mathcal{O}(- k \eta) \big] \;  \bigg) \,, \quad \notag
\end{eqnarray}
which is used to derive Eq.~(\ref{inflation_Iksq_LT}) in the main text.

There are also two notable special cases for the above formula. In the case where the environment is massless with $6 h_2 = m^2/H^2 = 0$ (so that $\mu = \frac{3}{2} i$) one finds instead
\begin{eqnarray} \label{Ik_massless}
I_{\bm{k}}(\eta) & = & \sqrt{ \frac{ \epsilon}{2} } \; f_1 \; \int_{-k \eta}^{\infty} \exd z'\; \frac{e^{+ 2 i z'} }{z'} \left( \frac{1}{z'} - i \right) \notag \\
& = & \sqrt{ \frac{ \epsilon}{2} } \; f_1 \; \left[ \frac{e^{- 2 i k \eta}}{- k \eta} - \pi - i \mathrm{Ei}(-2 i k\eta) \right] \quad (\mathrm{massless}), \qquad
\end{eqnarray}
where $\mathrm{Ei}$ is the exponential integral function. Furthermore, for the case of a conformal environment with $6 h_2 = m^2/H^2 = 2$ (so that $\mu = \frac{1}{2} i$) one finds
\begin{eqnarray}\label{eq:conformal}
I_{\bm{k}}(\eta) & = & - i \sqrt{ \frac{ \epsilon}{2} } \; f_1 \; \int_{-k \eta}^{\infty} \exd z'\; \frac{e^{+ 2 i z'} }{z'} \notag \\
& = & \sqrt{ \frac{ \epsilon}{2} } \; f_1 \; \left[ \pi + i \mathrm{Ei}(-2 i k\eta) \right] \quad (\mathrm{conformal}) \ . \qquad
\end{eqnarray}

\subsection{Ekpyrotic case for $\epsilon = \frac{3}{2}$}
\label{App:Ik_ekpy}

Using the mode functions in the contracting phase with $\epsilon = \frac{3}{2}$ yields
\begin{eqnarray}
I_{\bm{k}}(\eta) = - i f_1 \sqrt{6} \; \sqrt{ \frac{\pi}{4} }\; e^{-  \tfrac{\pi}{2}\mu + i \tfrac{\pi}{4}} \int_{-k \eta}^{\infty} \exd z' \;  \frac{ e^{i z'} H^{(1)}_{i \mu}(z') }{ \sqrt{ z' }  } \left( 1 + \frac{3i}{z'} - \frac{3}{(z')^2} \right)\,,
\end{eqnarray}
where we use the system mode functions from Eq.~(\ref{pmodes_5}) and otherwise the form of $I_{\bm{k}}(\eta)$ remains the same as in Appendix \ref{App:Ik_inflation}. The primitive of the above $z'$ integral is given by
\begin{eqnarray}
\int \exd z' \; \frac{ e^{ i z'} H_{i \mu}^{(1)}(z') }{\sqrt{z'}} = - \frac{ 2 i e^{ + \tfrac{\pi\mu}{2} }  \sqrt{z'} }{ \pi } \; \bigg( \; h_{\mu}(z') \; + \; h_{-\mu}(z') \; \bigg)\,,
\end{eqnarray}
where we define for convenience
\begin{equation}
h_{\mu}(z') := g_{\mu}(z') + \frac{6\pi e^{\tfrac{\pi\mu}{2}} \mathrm{csch}(\pi\mu) e^{i z'}}{ 9 + 4 \mu^2 } \bigg[ - \left( 1 + \sfrac{i}{z'} \right) J_{1 + i \mu}(z') + \left( - i + \sfrac{ \tfrac{3}{2} + i \mu }{z'} \left( 1 + \sfrac{i}{z'} \right)  \right) J_{i \mu}(z')  \bigg]\,,
\end{equation}
where $J_{i\mu}(z')$ is a Bessel function of the first kind and the function $g_{\mu}(z')$ defined above in Eq.~(\ref{gmu_def_App}). Evaluating at the end points of integration we find that
\begin{eqnarray}
I_{\bm{k}}(\eta) = - i \sqrt{ 6 }  f_1 \; \bigg( \; \frac{\pi \; \mathrm{sech}(\pi\mu) }{\sqrt{2}} - e^{- i \tfrac{\pi}{4}} \sqrt{ \frac{ - k \eta}{\pi}  } \; \big[ \; h_{\mu}(- k \eta) + h_{-\mu}(- k\eta) \; \big] \; \bigg) 
\end{eqnarray}
as quoted in the main text. Note that in the special case of $\mu = \frac{3}{2}i$ the above we find 
\begin{eqnarray} 
I_{\bm{k}}(\eta) & = & \sqrt{3} f_1 \int_{-k \eta}^{\infty} \exd z' \; e^{2i z'} \left( -\frac{3}{(z')^4}+\frac{6 i}{(z')^3}+\frac{4}{(z')^2}-\frac{i}{z'} \right) \notag \\
& = & \sqrt{3} f_1 \bigg[ \pi + i \mathrm{Ei}(- 2 i k \eta ) - e^{- 2 i k \eta} \left( \frac{1}{(- k \eta)^3} + \frac{2i}{k \eta} \right) \bigg] \qquad \qquad (\mu = \tfrac{3}{2} i ) \label{massless_ekpy_Ik}
\end{eqnarray}
which results in the scaling Eq.~(\ref{massless_ekpy_body}).

\subsection{Ekpyrotic case for $\epsilon > \frac{3}{2}$}
\label{App:caseiii}

In the case of case (iii) contraction we have $\epsilon > \frac{3}{2}$ with the additional condition $\Delta = \frac{3}{2} - \epsilon$ -- using Eq.~(\ref{spectral}) this enforces
\begin{equation} \label{caseiii_const}
 f_1^2 - \frac{1}{2} f_2 - \frac{1}{2} \left( 1 - \frac{3}{\epsilon} \right) h_2 = \frac{3}{2} - \epsilon \  .
\end{equation}
The system mode functions (\ref{pmodes_def}) consists of Hankel functions $\mathrm{p}_{\bm{k}}(\eta) \propto H_{\gamma - 1}(- k \eta)$ where
\begin{equation}
\gamma := \frac{\epsilon - 3}{2(\epsilon - 1)}
\end{equation}
while the environment mode functions (\ref{xmodes_def}) consist of Hankel functions $\mathrm{x}_{\bm{k}}(\eta) \propto H_{i\mu}(- k \eta)$ where
\begin{equation} \label{nu_def_app_pre}
\mu \ = \ \sqrt{ \frac{- 9 + 30 \epsilon - 17 \epsilon^2 - 16 f_1^2 \epsilon}{4(\epsilon-1)^2} } \ = : \ - i \nu 
\end{equation}
where we have used (\ref{caseiii_const}) to relate $24 h_2 - 8 \epsilon (f_2 + h_2) = 24 \epsilon - 16 f_1^2 \epsilon - 16 \epsilon^2$. Note that we've usefully defined $\nu = i \mu$ in the above since the parameter $\nu$ is purely real for $\epsilon > \frac{3}{2}$. 

The perturbative integral $I_{\bm{k}}(\eta)$ here has the form 
\begin{eqnarray} \label{Ik_iii}
I_{\bm{k}}(\eta) = \frac{i\pi}{4} \cdot \frac{ \sqrt{2 \epsilon} f_1}{\epsilon - 1} \cdot e^{ i \tfrac{\pi}{2} ( \gamma + \nu ) } \int_{-k \eta}^{\infty} \exd z' \;  H^{(1)}_{\gamma - 1}(z') H^{(1)}_{\nu}(z') \ ,
\end{eqnarray}
where since we work in perturbation theory, we have to leading-order in the coupling
\begin{equation} \label{nu_def_app}
\nu \ \simeq \ \sqrt{ \frac{9 - 30 \epsilon + 17 \epsilon^2 }{4(\epsilon-1)^2} }  + \mathcal{O}(f_1^2) \ .
\end{equation}
It is possible to find a closed-form expression for (\ref{Ik_iii}) in terms of hypergeometric functions, but we simply quote here its late-time limit which is 
\begin{eqnarray}
I_{\bm{k}}(\eta) & \simeq & i \frac{\sqrt{\epsilon} f_1}{\epsilon - 1} \bigg( \frac{\pi}{\sqrt{2} [ \cos(\pi \nu) - \cos(\pi \gamma) ]} \\
& \ &  + \ \frac{i - \cot(\pi \gamma)}{\sqrt{2} \Gamma(\gamma)}\bigg[ \frac{e^{+\tfrac{i \pi }{2} (\gamma - \nu)} \Gamma(-\nu) }{\gamma + \nu} \left( \frac{- k \eta}{2} \right)^{\gamma + \nu}  + \frac{e^{+\tfrac{i \pi }{2} (\gamma +  \nu)} \Gamma(\nu) }{\gamma - \nu} \left( \frac{- k \eta}{2} \right)^{\gamma - \nu}   \bigg] \bigg) + \ldots \notag
\end{eqnarray}
for $- k \eta \ll 1$. Since we have here $- \frac{3}{2} < \gamma < \frac{1}{2}$ as well as $\frac{3}{2} < \nu < \frac{3 \sqrt{2}}{2}$ for $\epsilon > \frac{3}{2}$, this means that the dominant term in the above is given by
\begin{eqnarray} \label{Ik_iii_App}
I_{\bm{k}}(\eta) & \simeq & i \; \frac{\sqrt{\epsilon} f_1}{\epsilon - 1} \; \cdot \; \frac{e^{+\tfrac{i \pi }{2} (\gamma +  \nu)} \big[ i - \cot(\pi \gamma) \big] \Gamma(\nu) }{\sqrt{2} (\gamma - \nu)\Gamma(\gamma)}\; \left( \frac{- k \eta}{2} \right)^{\gamma - \nu}  
\end{eqnarray}
which results in formula (\ref{Iksq_iii}) in the main text. 

\bibliographystyle{JHEP}
\bibliography{biblio}

\end{document}

%% file: newcommands.tex
\let\oldsqrt\sqrt
\def\sqrt{\mathpalette\DHLhksqrt}
\def\DHLhksqrt#1#2{%
\setbox0=\hbox{$#1\oldsqrt{#2\,}$}\dimen0=\ht0
\advance\dimen0-0.2\ht0
\setbox2=\hbox{\vrule height\ht0 depth -\dimen0}%
{\box0\lower0.4pt\box2}}

\newcommand{\ee}{e}

\newcommand{\boldmathsymbol}[1]{{\ensuremath{\boldsymbol{#1}}}}

\newcommand{\bmk}{\boldmathsymbol{k}}

\newcommand{\mpl}{m_\usssPl}

\newcommand{\beq}{\begin{equation}}
\newcommand{\eeq}{\end{equation}}
\newcommand{\bea}{\begin{equation}\begin{aligned}}
\newcommand{\eea}{\end{aligned}\end{equation}}

\newlength{\wsingfig}
\newlength{\wdblefig}
\newlength{\wquadfig}
\newlength{\wtriplefig}
